\documentclass[ALICE,manyauthors]{cernphprep}

\usepackage[comma,square,numbers,sort&compress]{natbib}
\usepackage{hyperref}
\usepackage{lineno}
\usepackage{url}
\usepackage{graphicx}
\usepackage{caption}
\usepackage{graphics}
\usepackage{multirow}
\usepackage{array}
\newcolumntype{P}[1]{>{\centering\arraybackslash}p{#1}}
\newcolumntype{M}[1]{>{\centering\arraybackslash}m{#1}}
\newcolumntype{H}{>{\setbox0=\hbox\bgroup}c<{\egroup}@{}}
\usepackage{hyperref}
\usepackage{amsmath}
\usepackage{breqn}
\usepackage{afterpage}
\usepackage{appendix}
\usepackage{pdfpages}
\begin{document}
\begin{titlepage}

\PHyear{2018}
\PHnumber{181}      
\PHdate{24 June}  

\title{Measurement of dielectron production in central Pb--Pb collisions \newline at $\mathbf{\sqrt{{\textit{s}}_{\mathrm{NN}}}}$ = 2.76 TeV}
\ShortTitle{Dielectrons in central Pb--Pb collisions}   

\Collaboration{ALICE Collaboration\thanks{See Appendix~\ref{app:collab} for the list of collaboration members}}
\ShortAuthor{ALICE Collaboration} 

\begin{abstract}

The first measurement of dielectron ($\mathrm{e}^{+}\mathrm{e}^{-}$) production in central (0$-$10$\%$) Pb--Pb collisions at $\mathbf{\sqrt{{\textit{s}}_{\mathrm{NN}}}}$ = 2.76 TeV at the LHC is presented. 
The dielectron invariant-mass spectrum is compared to the expected contributions from hadron decays in the invariant-mass range $0 < m_{\mathrm{ee}}<3.5\ \mathrm{GeV}/\textit{c}^{2}$.
The ratio of data and the cocktail of hadronic contributions without vacuum $\rho^{0}$ is measured in the invariant-mass range $0.15 < m_{\mathrm{ee}}<0.7\ \mathrm{GeV}/\textit{c}^{2}$, where an excess of dielectrons is observed in other experiments, and its value is $1.40 \pm 0.28\ (\mathrm{stat.}) \pm 0.08\ (\mathrm{syst.}) \pm 0.27\ (\mathrm{cocktail})$. 
The dielectron spectrum measured in the invariant mass range $0 < m_{\mathrm{ee}}<1\ \mathrm{GeV}/\textit{c}^{2}$ is consistent with the predictions from two theoretical model calculations that include thermal dielectron production from both partonic and hadronic phases with in-medium broadened $\rho^{0}$ meson. 
The fraction of direct virtual photons over inclusive virtual photons is extracted for dielectron pairs with invariant mass $0.1 < m_{\mathrm{ee}}<0.3\ \mathrm{GeV}/\textit{c}^{2}$, and in the transverse-momentum intervals $1<p_{\mathrm{T}, \mathrm{ee}}<2\ \mathrm{GeV}/\textit{c}$ and $2<p_{\mathrm{T}, \mathrm{ee}}<4\ \mathrm{GeV}/\textit{c}$. The measured fraction of virtual direct photons is consistent with the measurement of real direct photons by ALICE and with the expectations from previous dielectron measurements at RHIC within the experimental uncertainties.

\end{abstract}
\end{titlepage}
\setcounter{page}{2}

\section{Introduction}
\label{sec:Introduction}

The primary goal of studying ultra-relativistic heavy-ion collisions is to investigate the properties of the Quark--Gluon Plasma (QGP), a phase of matter created at extreme conditions of high temperature and energy density in which quarks and gluons are deconfined \cite{QCDequationOfState,chiralDeconfinementQCD}. 
\newline 
The space--time evolution of the collisions can be described by relativistic viscous hydrodynamics \cite{relativisticHydro} and hadronic transport models, which provide theoretical descriptions of the system in the partonic and hadronic phases respectively, taking into account many-body properties of Quantum-Chromodynamics (QCD). 
Photons and dileptons, i.e. lepton--antilepton pairs produced by internal conversion of virtual photons, are unique tools to study the space--time evolution of the hot and dense matter created in ultra-relativistic heavy-ion collisions. 
These electromagnetic probes are produced continuously by a variety of sources during the entire history of the collision and they traverse the medium with negligible final state interaction, thus carrying undistorted information on their production source \cite{directPhotonsHIcollisions}. 
\newline
Photons emitted by the thermal system, both in the partonic and in the hadronic phases, and those produced in the initial hard parton-parton scattering are referred to as direct photons. Thermal radiation carries information on the average temperature of the system, while photons from the initial scattering provide a test for perturbative QCD (pQCD) calculations. 
The largest background contribution to the direct photon measurements is represented by photons originating from hadron decays in the late stages of the collision. This physical background is usually described by the so-called `hadronic cocktail', obtained from simulations which incorporate the description of detector effects. Hadrons are generated according to their measured relative abundances and their decays are simulated based on the measured branching ratios \cite{PDG}. Their momentum distributions are parametrized from the measured spectra, or obtained using $m_{\mathrm{T}}$-scaling starting from some reference distribution when the measurement is not available.
\newline
One of the main motivations of measuring direct photons is to study the different contributions of the direct photon spectrum, with particular interest to its thermal component, which dominates the spectrum in the low-$p_{\mathrm{T}}$ region ($p_{\mathrm{T}} \lesssim 3\ \mathrm{GeV}/\textit{c}$). 
\newline
The direct photon measurements in Au--Au collisions at $\mathbf{\sqrt{{\textit{s}}_{\mathrm{NN}}}}$ = 200 GeV at RHIC \cite{directPhotonsAuAuPHENIX,enhancedDirectPhotonsAuAuPHENIX,directPhotonsAuAuSTAR} and in Pb--Pb collisions at $\mathbf{\sqrt{{\textit{s}}_{\mathrm{NN}}}}$ = 2.76 TeV at the LHC \cite{directPhotonsPbPbALICE} indicate a dominant contribution from early stages at high temperatures. This is in contradiction with the large direct-photon elliptic flow, measured by PHENIX at RHIC \cite{DirectPhotonsFlowPHENIX} and by ALICE at the LHC \cite{DirectPhotonsFlowALICE}, whose magnitude is comparable to the elliptic flow of all charged hadrons, suggesting a larger contribution from the late stages of the collisions. Theoretical model calculations based on relativistic hydrodynamics \cite{ellipticFlowThermalPhotons} fail in the simultaneous description of the direct photon spectrum and flow measured at  RHIC, while a smaller tension is observed at the LHC. This `direct photon puzzle' is still under study at the present time.
\newline
The study of real direct photons in the low-$p_{\mathrm{T}}$ region is challenging due to the large contribution of background photons originating from hadron decays. The dominant contribution is from $\pi^{0}$ decays (about 90$\%$). 
The main advantage of dielectrons over real photons is that this contribution can be significantly reduced by measuring virtual direct photons in the invariant-mass range above the $\pi^{0}$ mass. 
The fraction of virtual direct photons over inclusive virtual photons, measured in the kinematic range of quasi-real virtual photons ($p_{\mathrm{T, ee}} \gg m_{\mathrm{ee}}$), is expected to be identical to that of real photons in the zero mass limit. Therefore, the measurement of direct $\mathrm{e}^{+}\mathrm{e}^{-}$ pairs represents an independent and complementary measurement to that of real direct photons. 
\newline
Different intervals of the dielectron invariant-mass spectrum are sensitive to different stages of the collision and their related physical processes. 
\newline
The low-mass region of the dielectron spectrum ($m_{\mathrm{ee}} < m_{\phi}$) contains contributions from low-mass vector meson decays and thermal radiation from partonic and hadronic phases. In this mass range, the dielectron production is largely mediated by the $\rho^{0}$, $\omega$ and $\phi$ mesons. 
Among these, the $\rho^{0}$ is the most relevant source, due to its strong coupling to the $\pi^{+}\pi^{-}$ channel ($\pi^{+}\pi^{-} \rightarrow \rho^{0} \rightarrow \gamma^{*} \rightarrow \mathrm{e}^{+}\mathrm{e}^{-}$) and its lifetime of only 1.3 fm, making it subject to regeneration in the longer-lived hadronic gas phase.
A significant broadening of the electromagnetic spectral function of the $\rho^{0}$ meson in the hot hadron gas is produced as a consequence of many-body properties of hadron interactions in the hot hadron gas phase.
These are connected to the partial chiral symmetry restoration which is expected at temperatures close to the phase boundary with pseudo-critical temperature $T_{\mathrm{pc}} = 154 \pm 9$ MeV \cite{rhoMeltingCSR, pseudocriticalTemperature, equationOfStateQCD}. 
\newline
The intermediate-mass region of the dielectron spectrum ($m_{\phi} < m_{\mathrm{ee}} <m_{\mathrm{J}/\psi}$) is sensitive to direct thermal radiation from the partonic phase (QGP) and to correlated semileptonic charm and beauty decays.
In this mass range, the shape of the invariant-mass distribution gives information on the production mechanism of heavy quarks and on their angular correlations.  
The dielectrons in the high-mass region ($m_{\mathrm{ee}} > m_{\mathrm{J}/\psi}$) contain contributions from semileptonic heavy-flavor decays, heavy quarkonia, and Drell-Yan processes. Their initial production is described by perturbative QCD calculations and modified nuclear parton distribution functions and their final yield can be influenced by the presence of a deconfined state.
\newline
The interest in studying the dilepton continuum dates back to the 1970s, triggered by the experimental detection of the Drell-Yan process ($\mathrm{q\overline{q}} \rightarrow \gamma^{*} \rightarrow \mathrm{l}^{+}\mathrm{l}^{-}$) \cite{DrellYan} and the discovery of the J/$\psi$ \cite{JParticle,Psi}. 
The study of dielectrons was pioneered by the first generation of experiments at the CERN Super-Proton Synchrotron (SPS), which started its heavy-ion experimental program in 1986 using fixed target experiments.
The Helios-2 and NA38 experiments found an indication for an anomalous dilepton excess \cite{jPsiPsiPrimeNA38}, which was then confirmed by the second generation of SPS experiments. This experiment measured the dielectron spectra in different collision systems and center-of-mass energies. While the dielectron spectra measured in proton$-$nucleus collisions at $E_{\mathrm{lab}}$ = 450 GeV were compatible with the known hadronic sources, the spectra measured in nucleus$-$nucleus collisions at $E_{\mathrm{lab}}$ = 200 $\textit{A}$ GeV showed a clear excess in the dielectron production compared to the expected contributions from hadronic sources, especially in the low-mass region, below the $\rho^{0}$ mass \cite{dileptonEnhancementCERES}. 
The role of the $\rho^{0}$ meson as a mediator in the processes which produce thermal dileptons in the hot hadronic gas had already been highlighted a few years before and most of the theoretical models predicted an enhancement of the $\rho^{0}$, due to regeneration via $\pi^{+}\pi^{-} \rightarrow \rho^{0}$, but no in-medium effects. 
Theoretical models used to describe CERES data indicated two possible scenarios: a dropping of the $\rho^{0}$ pole mass or a broadening of its width \cite{DroppingMass,RhoBroadening,dielectronProductionPbAuCERES}.
Unfortunately, the mass resolution and statistical precision of the first CERES data did not provide a clear discrimination between these two possible scenarios, which both fitted the data similarly well. 
\newline
An enhanced dilepton production was also observed in the intermediate-mass region by Helios-3 for S--W with respect to p--W collisions at $E_{\mathrm{lab}}$ = 200 $\textit{A}$ GeV \cite{dimuonProductionpWSW,excessDimuonContinuumSW}. Theoretical arguments based on enhanced open-charm production and thermal radiation from the partonic phase were used to explain the excess, while experimentally this remained an open question.
The answers to the aforementioned ambiguities and unsolved questions came with NA60, a third-generation SPS experiment, specifically designed and built for dilepton measurements. NA60 measured the dimuon invariant-mass spectrum in In--In collisions at $E_{\mathrm{lab}}$ = 158 $\textit{A}$ GeV \cite{rhoSpectralFunctionNA60,electromagneticProbesInIn}. 
The mass resolution and the statistical precision of the measurement allowed for a detailed analysis of the $\rho^{0}$ spectral function leading to a clear conclusion regarding the origin of the excess observed in the low-mass region: the vacuum $\rho^{0}$ and the scenario based on the mass shift were ruled out, while the $\rho^{0}$-broadening was confirmed. 
The ambiguity of the first CERES data regarding the origin of the low-mass excess was solved with the new data in Pb--Au collisions at $E_{\mathrm{lab}}$ = 158 $\textit{A}$ GeV after the TPC upgrade \cite{dielectronPairsCentralPbAu}. Despite the fact that the calculations for both spectral functions described the enhancement reasonably well for masses below 0.7 GeV/$\textit{c}^{2}$, in the mass region between the $\rho^{0}$ and the $\phi$, the data clearly favored the broadening scenario over the dropping mass scenario, confirming the NA60 results.
\newline
The capability of the NA60 experiment to distinguish prompt muons, originating from the interaction vertex, and decay muons, coming from displaced vertices, allowed them to clearly establish the absence of any charm enhancement in the intermediate-mass region. The excess of dileptons previously seen in the intermediate-mass region, and confirmed by NA60, was solely attributed to prompt dimuons. 
The origin of this excess of prompt dilepton radiation dominating the intermediate-mass region of the dilepton spectrum was also investigated by studying the inverse slope parameter $T_{\mathrm{eff}}$ of the dimuon $p_{\mathrm{T}}$ spectrum as a function of mass \cite{thermalDimuonsNA60}. This study was the first indication that dilepton radiation in the intermediate-mass region originates from the partonic phase \cite{partonicRadiation,lowMassDimuonsRelativisticNuclearColl}.
\newline \indent 
The dilepton experimental program has been continued at higher energies at the Relativistic Heavy-Ion Collider (RHIC) by PHENIX and STAR. 
The first dielectron measurement in Au--Au collisions at a center-of-mass energy per nucleon pair $\mathbf{\sqrt{{\textit{s}}_{\mathrm{NN}}}}=200\ \mathrm{GeV}$ reported by PHENIX showed a very large excess in the low-mass region compared to the hadronic cocktail \cite{dielectronsAuAuPHENIX}. All models that successfully reproduced the SPS results failed to explain the PHENIX data \cite{dielectronsAuAuPHENIX, dileptonProductionAuAuPHSD}.
Later, the STAR Collaboration also measured the dielectron invariant-mass spectrum in Au--Au collisions at $\mathbf{\sqrt{{\textit{s}}_{\mathrm{NN}}}}=200\ \mathrm{GeV}$  \cite{dielectronsAuAuSTAR}. The enhancement observed by STAR  in the low-mass region is much smaller compared to PHENIX, and it is compatible with models that involve the broadening of the $\rho^{0}$ meson \cite{dielectronsAuAuSTAR}.   
The inconsistency in the measurements of the low-mass excess between PHENIX and STAR has been solved with the new data from PHENIX, collected in 2010 after the installation of the Hadron Blind Detector (HBD) \cite{dielectronProductionAuAuPHENIXnew}. The large excess seen in the 2004 data was not confirmed, and the measured enhancement in the low-mass region is now consistent with the STAR measurement, and with models predicting the $\rho^{0}$ broadening. 
\newline
The study of the dielectron spectrum in Pb--Pb collisions at the Large Hadron Collider (LHC), delivering higher collision energies, is interesting to further investigate the $\rho^{0}$ broadening effect near the phase transition to chiral symmetry restoration. 
Moreover, the study of the thermal dielectrons provides information on the temperature of the system created at unprecedented center-of-mass energies contributing to the characterization of its thermodynamic properties. 
The system created in Pb--Pb collisions at LHC energies has vanishing baryon chemical potential and hence the thermal component of the dielectron spectrum, including the effects of $\rho^{0}$ broadening, can be calculated using a lattice-QCD inspired approach based on a equation of state.
The dielectron measurement at the LHC energies is extremely challenging due to the very small signal-to-background ratio, that is of the order of $10^{-3}$ in the low-mass region, which requires very high precision in the estimate of the combinatorial background and a detailed knowledge of the hadronic contributions to the dielectron spectrum. 
\newline 
In this paper, the first measurement of dielectron production in central (0$-$10$\%$) Pb--Pb collisions at $\mathbf{\sqrt{{\textit{s}}_{\mathrm{NN}}}}$ = 2.76 TeV with ALICE at the LHC is reported. The data used in this analysis were recorded during the LHC heavy-ion run in the year 2011.
\newline  
The paper is organized as follows. Section~\ref{sec:TheALICEapparatus} contains a brief description of the ALICE apparatus. Section~\ref{sec:DataAnalysis} illustrates the analysis techniques, including the particle-identification methods, the photon-conversion rejection techniques, the background description and signal extraction.
Section~\ref{sec:Results} presents the results on dielectron production yields within the ALICE acceptance and the comparison of the measured spectrum to the expectations from known hadronic sources and theoretical calculations. Section~\ref{sec:Conclusions} presents our summary and outlook.

\section{The ALICE apparatus}
\label{sec:TheALICEapparatus}

ALICE (A Large Ion Collider Experiment) is a detector system at the LHC specifically dedicated to the study of heavy-ion collisions. The characteristics of the ALICE apparatus are described in detail in \cite{ALICE}. Electrons are measured in the ALICE central barrel using the Inner Tracking System (ITS) \cite{ITSalignment}, the Time Projection Chamber (TPC) \cite{TPC} and the Time-Of-Flight (TOF) \cite{TOF} system. 
The central barrel detectors are located within a large solenoidal magnet, providing a field $\textit{B}$ = 0.5 T parallel to the beam line ($\textit{z}$ axis of the ALICE reference frame).
The collision centrality is estimated using the measured charged-particle multiplicity in two scintillator hodoscopes (V0-A and V0-C detectors), covering the pseudorapidity regions -3.7 $< \eta <$ -1.7 and 2.8 $< \eta <$ 5.1.   
\cite{VZEROPerformance}. The centrality is defined in terms of percentiles of the total hadronic cross section \cite{CentralityDetermination}.
\newline
The ITS consists of six cylindrical layers of silicon detectors, concentric and coaxial to the beam pipe, with a total pseudorapidity coverage $|\eta| < 1.2$. Three different technologies are used for this detector: the two innermost layers consist of Silicon Pixel Detectors (SPD), the two central layers of Silicon Drift Detectors (SDD) and the two outermost layers of double-sided Silicon Strip Detectors (SSD). The detector radii range from 3.9 cm for the innermost layer up to 43 cm for the outermost layer. The ITS is used in the determination of the primary and secondary vertices, and in the track reconstruction in the vicinity of the collision point. 
The last four layers of the ITS detector have particle-identification capabilities, via specific energy loss (d$\textit{E}$/d$\textit{x}$) measurements in the silicon detectors, which are complementary to the Particle IDentification (PID) signals from other detectors. 
The TPC is the main tracking device in the ALICE central barrel, with a pseudorapidity coverage $|\eta| < 0.9$. It is used for track finding and reconstruction, charged-particle momentum measurement via their curvature radius in the magnetic field and for particle identification via the measurement of the particle's specific energy loss in the TPC gas. 
The TPC is cylindrical in shape, coaxial with the beam pipe, with an active gas volume ranging from about 85 cm to 250 cm in the radial direction, and a length of 510 cm in the beam direction. The TPC volume is divided into two symmetric parts by a thin high-voltage electrode at 100 kV, parallel to and equidistant from the two endcaps, which is used to create a highly uniform electrostatic field in the two drift regions of $\sim250$ cm length inside the field cage. 
The gas mixture used,  $90\%\ \mathrm{Ne}$ and $10\%\ \mathrm{CO}_{2}$, is characterized by low diffusion, low-$\textit{Z}$, and large ion mobility. These requirements are needed for a good momentum and PID resolution, and to guarantee the highest possible data acquisition rate.  
The TOF detector is made of Multigap Resistive Plate Chambers (MRPC), with a pseudorapidity coverage $|\eta| < 0.9$ and a time resolution of $\sim 80$ ps. This detector is arranged in a modular structure with 18 blocks in azimuthal angle matching the TPC sectors. It has a cylindrical shape, covering polar angles between 45 degrees and 135 degrees over the full azimuth. This detector is used in the electron identification.
The determination of the event collision time needed to perform particle identification with the time-of-flight method can be provided on an event-by-event basis by the TOF detector itself or by the T0 detector \cite{ForwardDetectors}. The latter consists of two arrays of Cherenkov counters (T0C and T0A) positioned around the beam pipe, on both sides of the nominal interaction point. When both measurements of the start time are available, their weighted mean is used \cite{collisionTime}.

\section{Data analysis}
\label{sec:DataAnalysis}

\subsection{Data sample, event and track selection}
\label{subsec:EventAndTrackSelection}

The data sample used in this analysis was recorded with ALICE in 2011 during the LHC Pb--Pb run at $\mathbf{\sqrt{{\textit{s}}_{\mathrm{NN}}}}=2.76\ \mathrm{TeV}$ using a trigger on central collisions requiring a minimum number of charged tracks measured in the V0-A and V0-C detectors \cite{ForwardDetectors}. The measured multiplicity in the V0 detectors is also used for the offline selection of the centrality interval 0$-$10$\%$.
Two opposite magnetic field polarities were applied during the data taking and analysed separately due to geometrical asymmetries of the experimental apparatus (see Sec.~\ref{subsec:BackgroundDescriptionAndSignalExtraction}).
The pile-up from beam-gas interactions, collisions with de-bunched ions or with mechanical structures of the machine is reduced to a negligible level by rejecting events with multiple vertices identified with the SPD.
In order to keep the conditions of the detectors as uniform as possible, avoid edge effects and reject residual parasitic collisions, the coordinate of the primary vertex along the beam axis is required to be within 10 cm from the nominal interaction point of ALICE. The total number of events collected for the centrality class 0$-$10$\%$ is approximately 20 million. 
\newline   
Primary track candidates with transverse momentum $0.4 < p_{\mathrm{T}}< 5$ GeV/$\textit{c}$ and $|\eta|<0.8$ are selected from the reconstructed tracks in the ITS and TPC by applying some quality requirements. More specifically, tracks are required to have a minimum number of reconstructed space points in the TPC ($N_{\mathrm{cls}}^{\mathrm{TPC}}>70$) and in the ITS ($N_{\mathrm{cls}}^{\mathrm{ITS}}>4$), with one in the first SPD layer. The latter requirement is used to suppress the contribution of electrons from photon conversion in the detector material produced at radial distances larger than that of the first ITS layer. 
A minimum quality of the track fit is also required, expressed by $\chi^{2}/\mathrm{ndf}<4$ and a ratio of the number of reconstructed TPC clusters over the number of findable TPC clusters (accounting for track length, spatial location, and momentum) larger than 60$\%$. 
The contribution from secondary tracks is reduced using the distance of closest approach (DCA) to the primary vertex. 
A parametrized $p_{\mathrm{T}}$-dependent selection is applied to the transverse distance of closest approach: $\mathrm{DCA}^{\mathrm{max}}_{\mathrm{xy}} (p_{\mathrm{T}})= 0.005 + 0.01 / p_{\mathrm{T}}^{1.3}$ cm, with $p_{\mathrm{T}}$ in GeV/$\textit{c}$. The longitudinal DCA is required to be smaller than 0.1 cm.   
Only tracks with an associated hit in the TOF are accepted in order to exploit its particle-identification capabilities. The TOF radial distance and the ALICE magnetic field determine a minimum $p_{\mathrm{T}}$ of approximately 0.3 GeV/$\textit{c}$. The lower threshold on $p_{\mathrm{T}}$ is set to a higher value ($p_{\mathrm{T}}>0.4$ GeV/$\textit{c}$) due to the large inefficiency of the TOF and to the very low TPC--TOF matching efficiency for $p_{\mathrm{T}}$ close to 0.3 GeV/$\textit{c}$ \cite{ALICEperformance}.

\subsection{Electron identification and hadron contamination}
\label{subsec:ElectronIdentificationAndHadronContamination}

\begin{figure}[!hbt]
         \centering 
 	        \includegraphics[height=5.5cm,width=7.5cm]{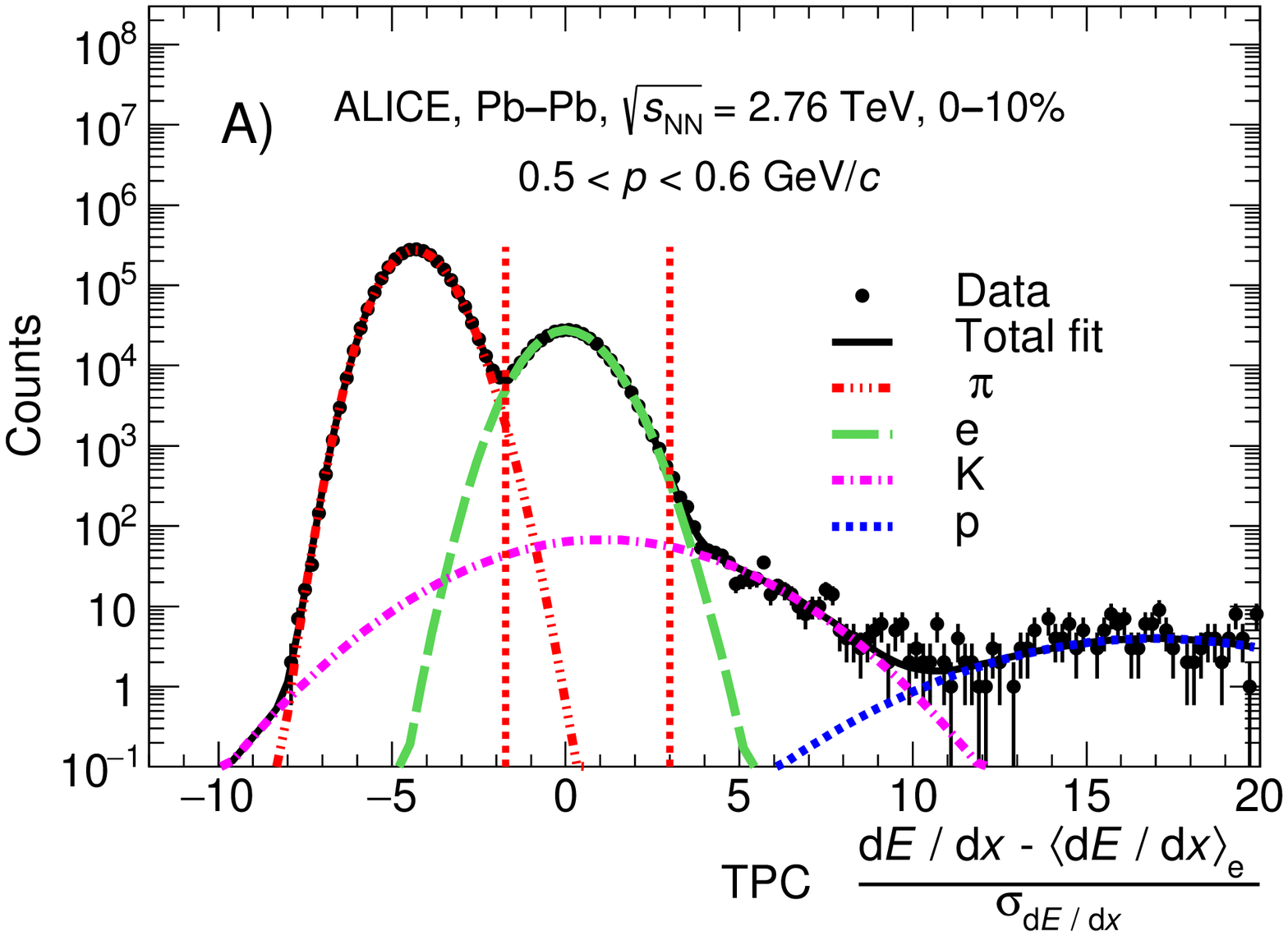}          
                 \includegraphics[height=5.5cm,width=7.5cm]{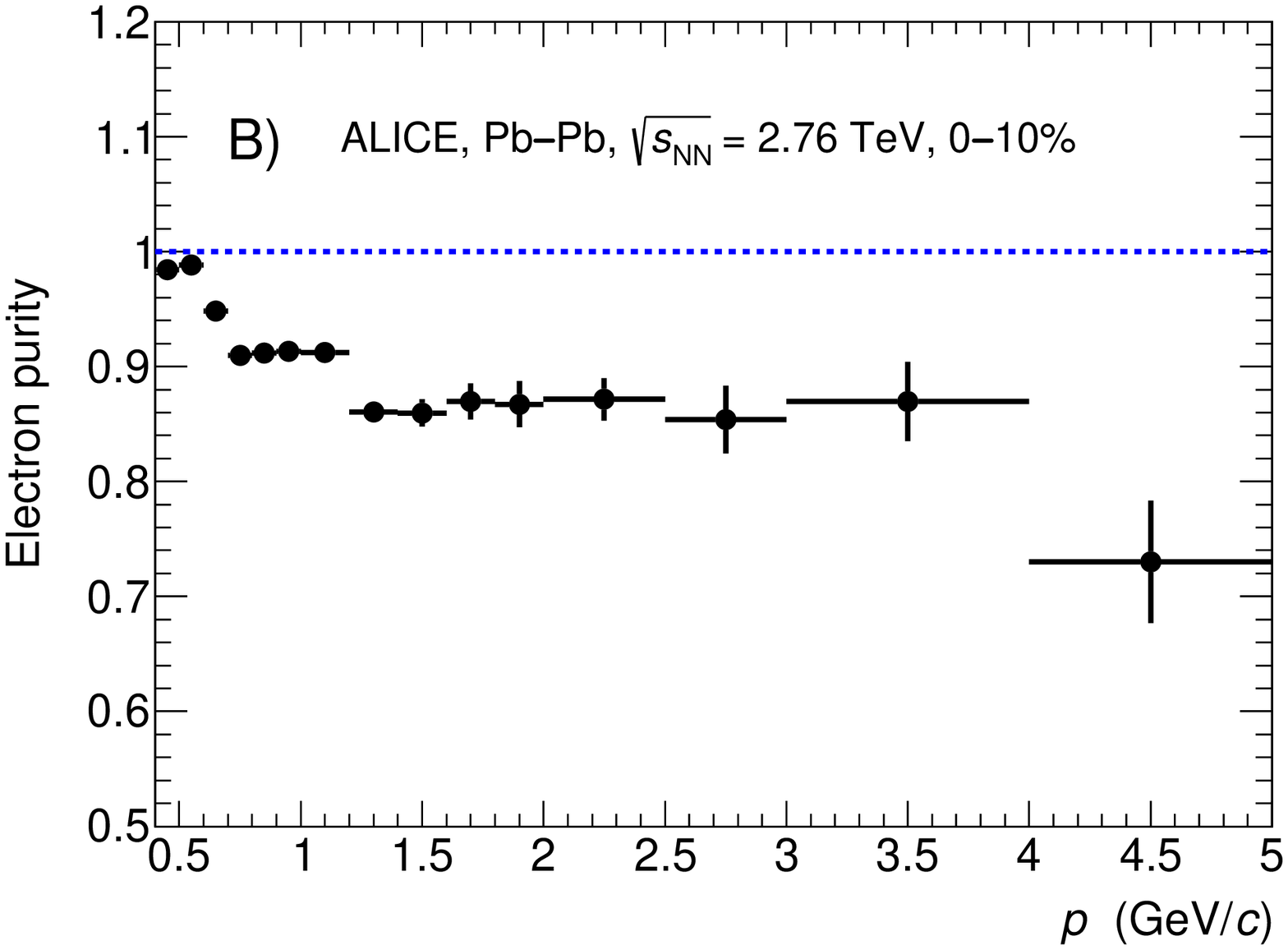}
         \caption{(colour online). (A) Distributions of the difference between the measured d$\textit{E}$/d$\textit{x}$ in the TPC and the expected average value for electrons (in units of $\sigma_{\mathrm{TPC}}$) after the selection of the electron candidates in the TOF ($|n\sigma_{\mathrm{TOF}}(\mathrm{e})|<3$) and in the ITS ($-5<n\sigma_{\mathrm{ITS}}(\mathrm{e})<1$). (B) Electron purity as a function of the track momentum ($p$) estimated from the multiple skewed Gaussian fit. }
 \label{fig:ePurity}
 \end{figure}

The particle-identification methods used in ALICE are illustrated in detail in \cite{ALICEperformance}.
To identify electrons, the time-of-flight of the particles is required to be consistent within 3$\sigma_{\mathrm{TOF}}$ with the expected value for the electron mass hypothesis, where $\sigma_{\mathrm{TOF}}$ is the PID resolution of the TOF detector. After this preselection, the specific energy loss measurements in the TPC and in the ITS are used for the selection of the electron candidates.   
In the TPC, the measured d$\textit{E}$/d$\textit{x}$ is required to be within an asymmetric range around the expected value for electrons: the upper limit is set at 3$\sigma_{\mathrm{TPC}}$, while a momentum-dependent lower limit is applied, parametrized as $-3\sigma_{\mathrm{TPC}} \cdot \mathrm{exp} (-p)$, where $p$ (GeV/$\textit{c}$) is the track momentum at the primary vertex. The latter is chosen due to the increasing overlap between the pion and electron bands at higher momenta.  
In order to improve the hadron rejection capabilities at low momentum, tracks are required to have a measured d$\textit{E}$/d$\textit{x}$ in the ITS in the range [-5$\sigma_{\mathrm{ITS}}$, 1$\sigma_{\mathrm{ITS}}$] around the expected value for electrons, being $\sigma_{\mathrm{ITS}}$ the d$\textit{E}$/d$\textit{x}$ resolution of the ITS.
The electron purity, defined as the fraction of electrons in the selected track sample, is estimated as a function of track momentum using a multiple skewed Gaussian fit of the TPC signal, after the selection of the electron candidates in the ITS and TOF.  
The skewed gaussian distribution is defined by $2g(x)\cdot \Phi(\alpha x)$, where $g(x)$ is the Gaussian function and $\Phi(x)=0.5 \cdot \left[ 1 + \mathrm{erf} (x/\sqrt{2}) \right]$. The skewness $\alpha$ is a free parameter of the fit, which varies for different particle species.
The difference between the measured d$\textit{E}$/d$\textit{x}$ in the TPC and the expected value for electrons, in units of $\sigma_{\mathrm{TPC}}$, for the momentum interval $0.5 < p < 0.6\ \mathrm{GeV}/\textit{c}$ is shown in Fig.~\ref{fig:ePurity}-A.  The electron purity, shown as a function of the track momentum in Fig.~\ref{fig:ePurity}-B, is estimated in each momentum interval by extracting the residual fraction of hadrons within the selected region. 
The overall purity is better than $90\%$ for low momenta ($p \lesssim 1.2\ \mathrm{GeV}/\textit{c}$) and better than $85\%$ at high momenta, except for the momentum interval $4 < p < 5\ \mathrm{GeV}/\textit{c}$ where the purity is approximately 70$\%$.

\begin{figure}[!hbt]
         \centering 
 	        \includegraphics[height=8cm,width=9.5cm]{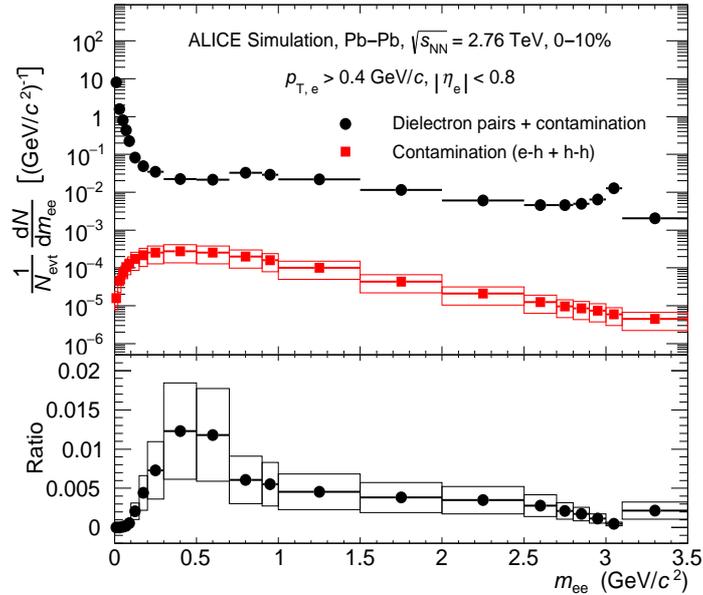}          
         \caption{(colour online). Invariant-mass distribution of correlated electron-hadron and hadron-hadron pairs, where hadrons are wrongly identified as electrons, in comparison with the invariant-mass spectrum of all pairs (dielectron signal pairs and `contaminated' pairs) obtained from MC simulations. In the bottom panel, the ratio between the invariant-mass spectrum of electron-hadron and hadron-hadron pairs to that of all pairs is shown. Boxes represent the systematic uncertainties on the hadron contamination. }
 \label{fig:HadronContaminationInMassSpectrum}
 \end{figure}

The impact of hadron contamination on the dielectron invariant-mass spectrum is studied using Monte Carlo (MC) simulations of central Pb--Pb events based on the HIJING generator \cite{HIJING}.   
The contribution of `contaminated pairs'  is given by the invariant-mass spectrum of correlated electron-hadron and hadron-hadron pairs, where hadrons are wrongly identified as electrons.
\newline
We consider as correlated pairs oppositely-charged particles produced in the decay of the same resonance or pairs from correlated $\mathrm{c}\overline{\mathrm{c}}$ and $\mathrm{b}\overline{\mathrm{b}}$ decays. 
In this MC study, the same level of hadron contamination estimated in data is reproduced in the simulation. This is done by randomly adding hadrons to a pure sample of electrons such that the hadron-to-electron ratio reproduces the same ratio estimated in real data, in each momentum interval, and for each hadron species.  
The input transverse-momentum distributions of dielectron sources in the MC simulation are corrected, using $p_{\mathrm{T}}$-dependent weights obtained from the measured spectra \cite{PizeroMeasurementPbPb,mtScaling,EtaToPiRatio,KToPiRatio}, in order to have the same relative particle abundances as in data.
The invariant-mass distribution of correlated electron-hadron and hadron-hadron pairs is shown in Fig.~\ref{fig:HadronContaminationInMassSpectrum} in comparison with the invariant-mass spectrum of all pairs (dielectron signal pairs and `contaminated' pairs) obtained from the simulation. The relative contribution of hadron contamination to the dielectron invariant-mass spectrum is maximum in the low-mass region ($0.2 < m_{\mathrm{ee}} < 0.7\ \mathrm{GeV}/\textit{c}^{2}$) with a peak around $m_{\mathrm{ee}} = 0.5\ \mathrm{GeV}/\textit{c}^{2}$ and it decreases at high mass to less than $1 \%$. The maximum value is less than $1.5 \%$ for the $p_{\mathrm{T}, \mathrm{ee}}$-integrated spectrum,  less than $1 \%$ for $1 < p_{\mathrm{T}, \mathrm{ee}} < 2\ \mathrm{GeV}/\textit{c}$ and less than $4 \%$ for $2 < p_{\mathrm{T}, \mathrm{ee}} < 4 \ \mathrm{GeV}/\textit{c}$. The estimated contribution from hadron contamination is subtracted from the measured dielectron spectrum in all transverse-momentum intervals. A systematic uncertainty of 50$\%$ is assumed on the relative contribution of hadron contamination to account for possible discrepancies between real data and simulations, e.g. missing sources of electrons and/or hadrons and branching ratios. The systematic uncertainty on the hadron contamination is added in quadrature to all other contributions  (see Sec.~\ref{subsec:SystematicUncertainties}).

\subsection{Conversion rejection methods}
\label{subsec:ConversionRejectionMethods}

\begin{figure}[!hbt]
         \centering 
                 \includegraphics[height=8cm,width=11cm]{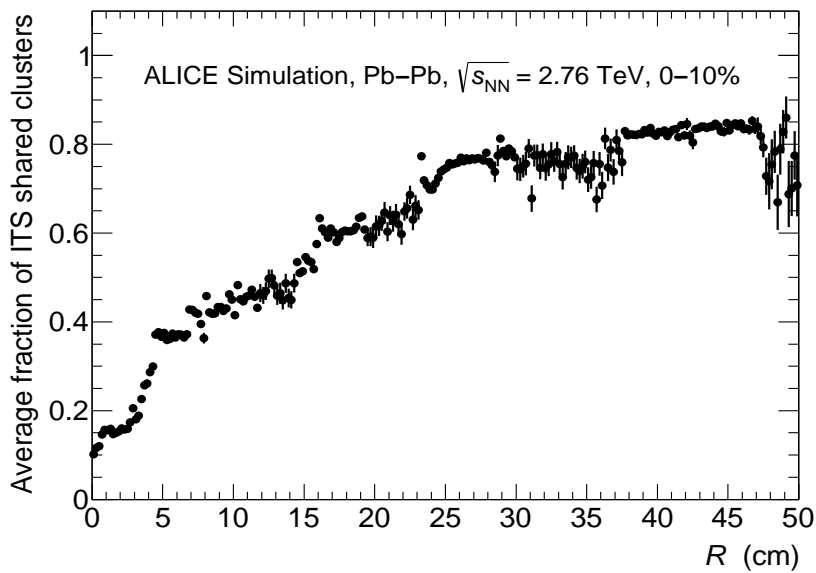}
         \caption{Average fraction of shared clusters in the ITS for secondary tracks as a function of the radial distance ($R$) between the true production vertex and the primary vertex obtained from MC simulations.}
 \label{fig:AverageFracSharedCls}
 \end{figure}

\begin{figure}[!hbt]
         \centering
                 \includegraphics[height=5.5cm,width=7.5cm]{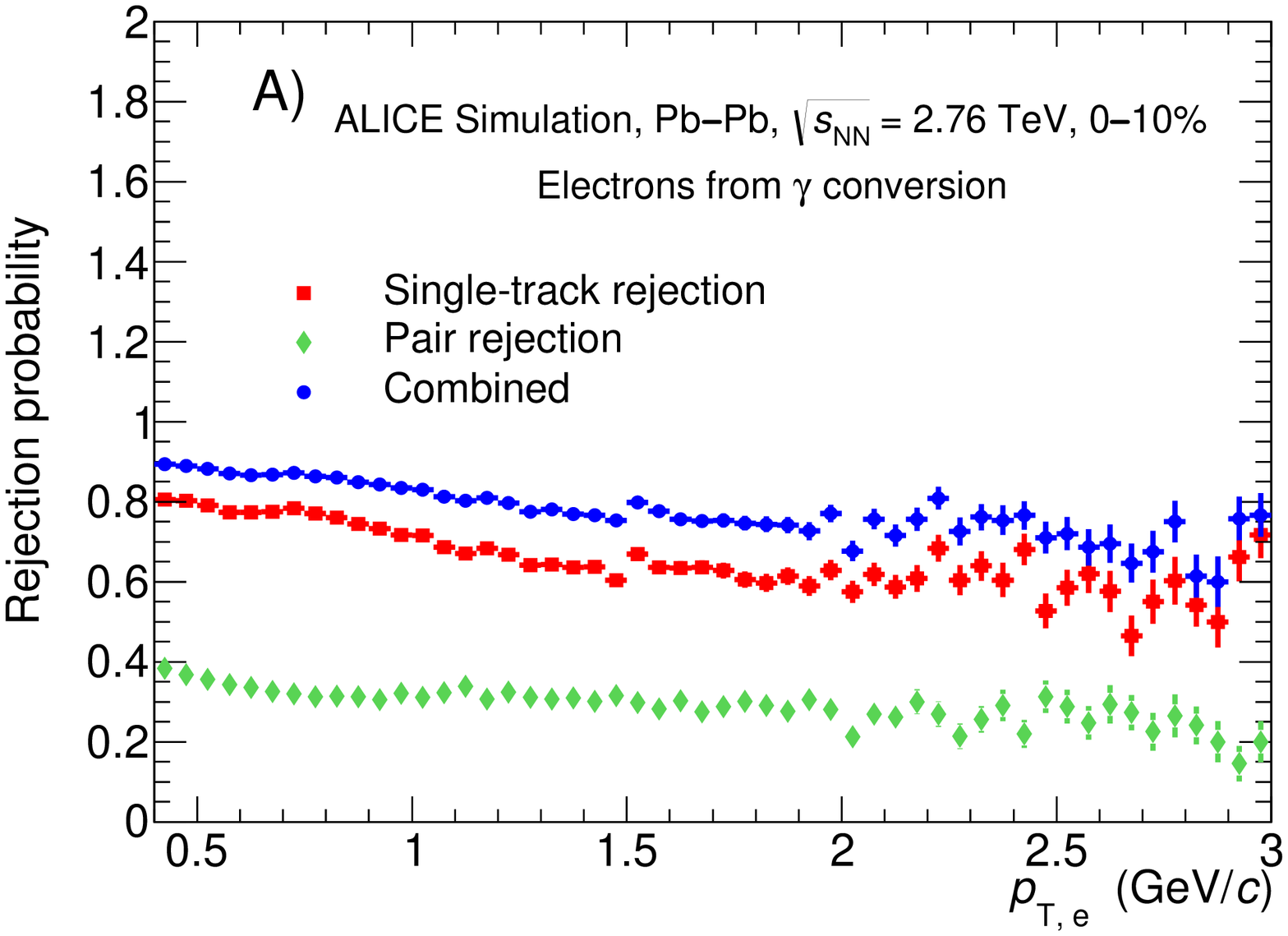}
                 \includegraphics[height=5.5cm,width=7.5cm]{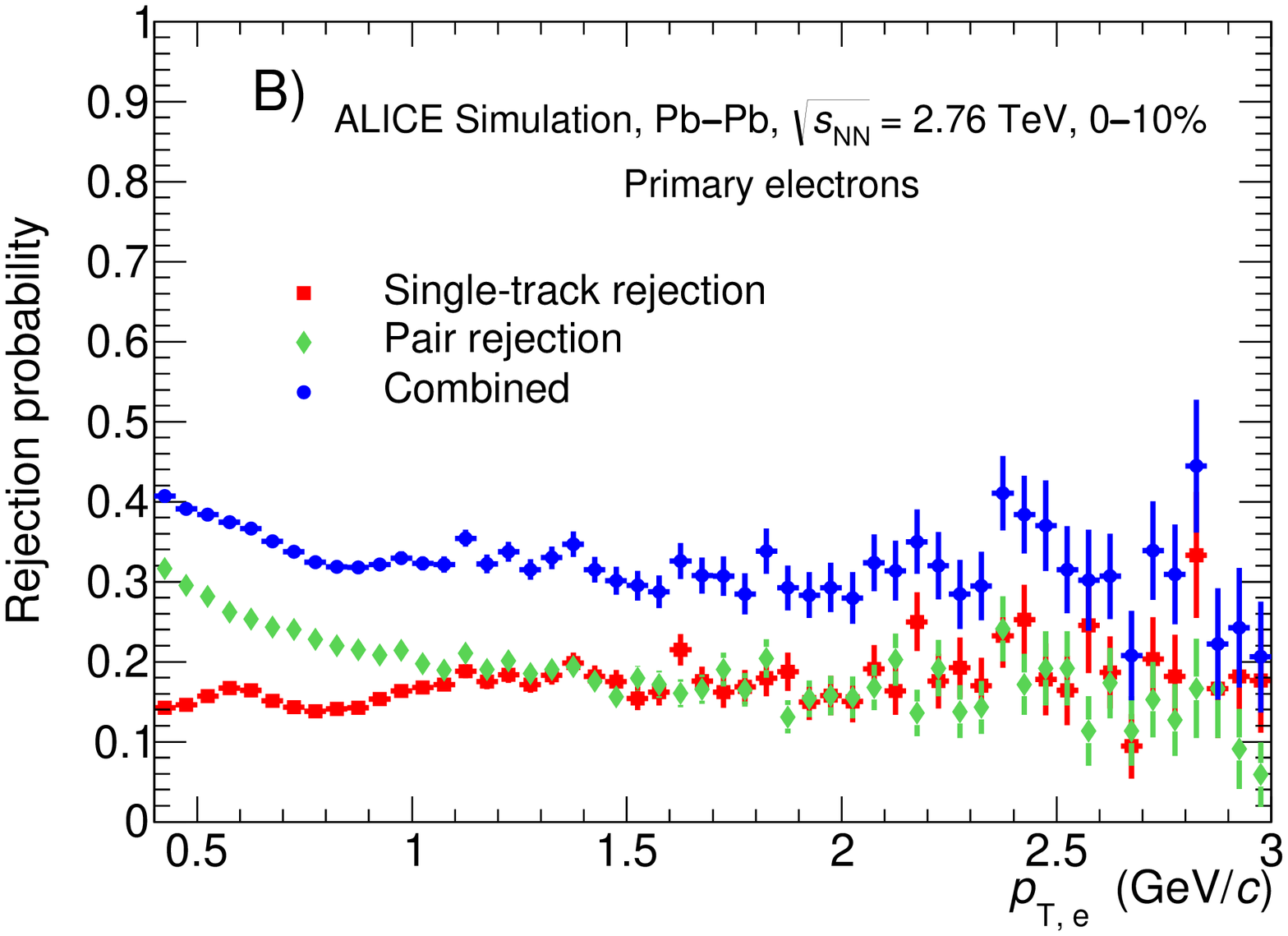}
         \caption{(colour online). (A) Rejection efficiency of conversion electrons as a function of $p_{\mathrm{T}, \mathrm{e}}$ using single-track rejection, pair rejection and using both methods. (B) Fraction of signal tracks rejected by the conversion rejection methods as a function of $p_{\mathrm{T}, \mathrm{e}}$. }
 \label{fig:ConversionRejectionEfficiency}
 \end{figure}

The largest contribution to the combinatorial background, originating from the sequential pairings between oppositely charged dielectron pairs (see Sec.~\ref{subsec:BackgroundDescriptionAndSignalExtraction}), is given by conversion electrons. These are more than 70$\%$ of all electrons in the selected track sample.
Since the signal-to-background ratio is of the order of $10^{-3}-10^{-2}$ in the low-mass region ($0.2 < m_{\mathrm{ee}} < 1\ \mathrm{GeV}/\textit{c}^{2}$), the suppression of the largest background contributor is a crucial aspect for the dielectron measurement.
Two methods for rejecting electrons from photon conversion in the detector material are applied. A single-track conversion rejection is complemented by a pair-rejection method which exploits the orientation of the plane spanned by the electron-positron pair produced by a photon conversion with respect to the magnetic field. 
\newline  \indent
Conversion rejection on a single-track level is based on the properties of the track reconstruction algorithm. All tracks are propagated down to the point of closest approach to the primary vertex. In the case of secondary tracks, this inward propagation results in wrong cluster associations in the ITS causing some distortions in the reconstructed track and in the measured momentum. Secondary tracks have shared clusters in the ITS with other primary tracks.
The average fraction of shared clusters for secondary tracks increases for larger distances between the true production vertex of conversion electrons and the primary vertex, as shown in Fig.~\ref{fig:AverageFracSharedCls}. A maximum 40$\%$ fraction of shared clusters in the ITS is allowed for all reconstructed tracks. This requirement represents the best compromise between the maximization of the photon conversion rejection and the minimization of signal loss. 
The single-track rejection method, which includes the requirement on the maximum fraction of shared ITS clusters, the requirement of having a cluster in the first ITS layer and the selection of tracks based on their DCA (see Sec.~\ref{subsec:EventAndTrackSelection}), is very efficient in rejecting single electrons from conversion in the case of a non-reconstructed partner. 
A large suppression of conversions is achieved by this technique, amounting to about $80\%$ ($60\%$) at low (high) $p_{\mathrm{T}, \mathrm{e}}$, with a loss of signal tracks of around $15\%$ approximately constant with $p_{\mathrm{T}, \mathrm{e}}$ (Fig.$~\ref{fig:ConversionRejectionEfficiency}$-B).

In the pair rejection method, electron candidates from photon conversions are tagged and rejected from the track sample if combined with any other electron they fulfill a tagging criterion, thus preventing them from contributing to the combinatorial background. 
Electron-positron pairs produced by photon conversion in the material have zero mass (neglecting the recoil momentum of the nucleus involved in the process) and, as a consequence, no intrinsic opening angle. These particles are bent only in the azimuthal direction by the magnetic field. The tracks produced by conversion electrons are propagated inward to the primary vertex by the tracking algorithm and their momenta are extrapolated at the point of closest approach to the primary vertex. This procedure creates an artificial opening angle (and mass), which becomes larger for increasing distances of the conversion point from the primary vertex.
Given a pair of particles, the vector connecting the ends of the momentum vectors of the two particles defines the orientation of their opening angle. The expected orientation of the opening angle of a conversion pair is given by: 

\begin{equation}
\vec{w}_{\mathrm{exp}} = \vec{p} \times \vec{z}
\end{equation}
 
where $\vec{p} = \vec{p_{1}} +  \vec{p_{2}}$ is the total momentum of the $\mathrm{e}^{+}\mathrm{e}^{-}$ pair, and $\vec{z}$ is the direction of the magnetic field (the $z$-axis). 
The measured orientation of the opening angle is given by: 

\begin{equation}
\vec{w}_{\mathrm{meas}} = \vec{p} \times \vec{u},
\end{equation}
 
where $\vec{u}$ is a unit vector perpendicular to the plane defined by the electron-positron pair ($\vec{u}=(\vec{p}_{1}\times\vec{p}_{2})/ |\vec{p}_{1}\times\vec{p}_{2}| $). A discriminating variable to identify conversion electrons is the angle ($\psi_{\mathrm{V}}$) between the expected and measured orientation of their opening angle:

\begin{equation}
\mathrm{cos} (\psi_{\mathrm{V}}) = \vec{w}_{\mathrm{exp}} \cdot \vec{w}_{\mathrm{meas}}
\end{equation}

\begin{figure}[!hbt]
         \centering
                 \includegraphics[height=5.5cm,width=7.5cm]{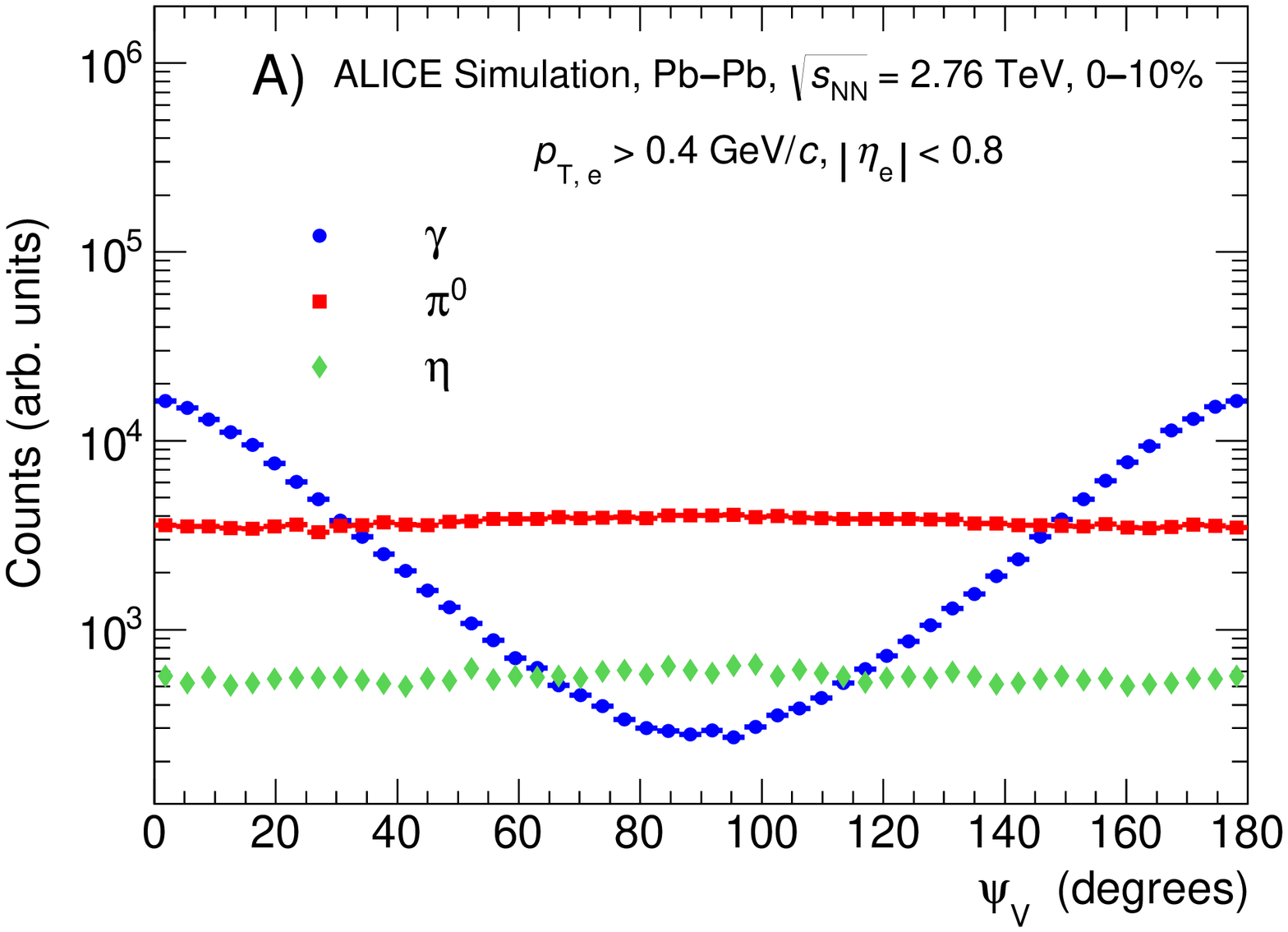}
                 \includegraphics[height=5.5cm,width=7.5cm]{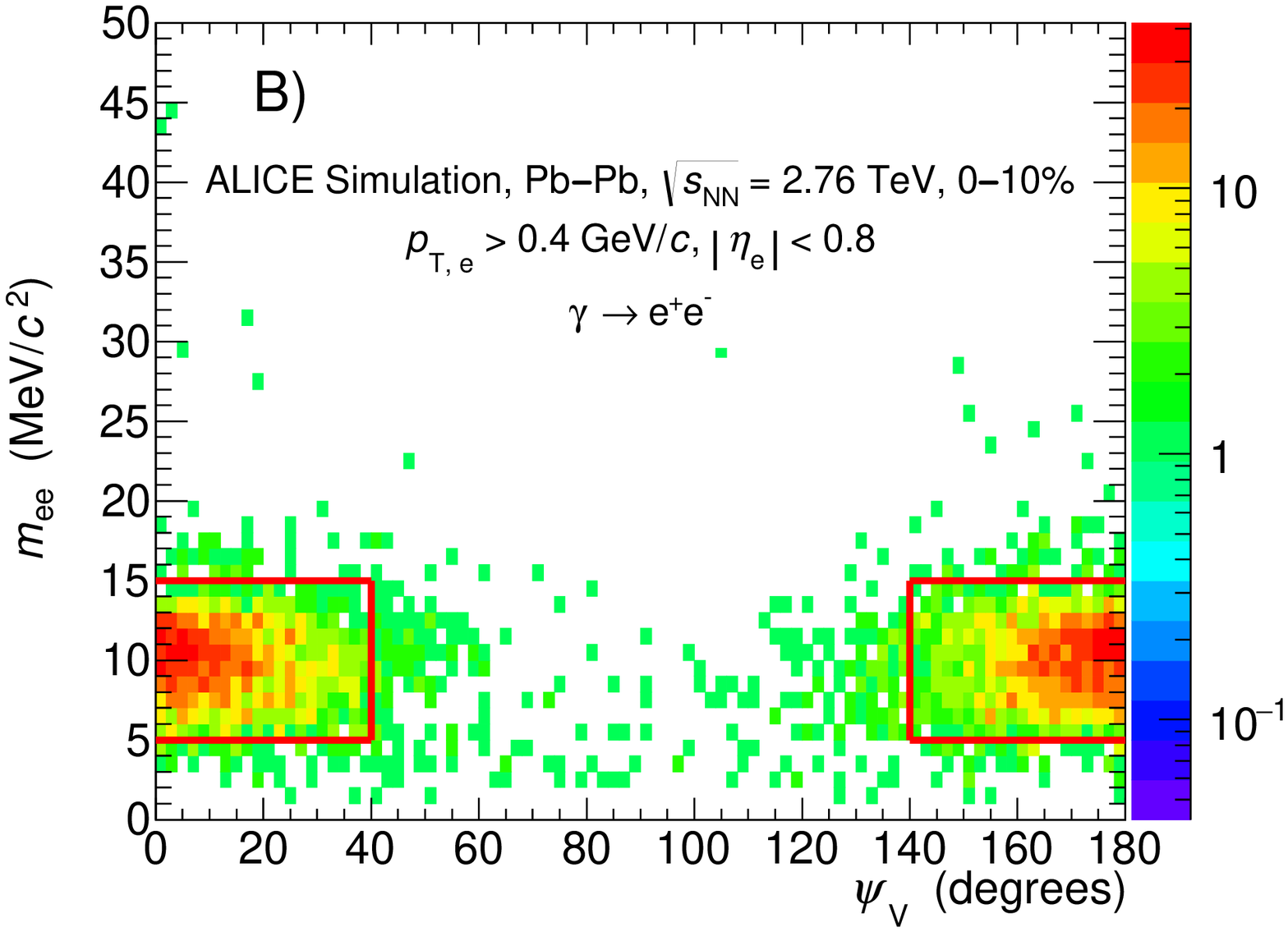}
         \caption{(colour online). (A) $\psi_{\mathrm{V}}$ distribution for correlated $\mathrm{e}^{+}\mathrm{e}^{-}$ pairs originating from different dielectron sources. (B) Correlation between $\psi_{\mathrm{V}}$ and  $m_{\mathrm{ee}}$ for electron-positron pairs produced by photon conversion in the detector material. Their tracks have passed the single-track conversion rejection. The regions used to select the conversion pair candidates are indicated by the red boxes.}
 \label{fig:PhiVDistribution}
 \end{figure}

Conversion pairs should have $\psi_{\mathrm{V}}$ = 0 or $\psi_{\mathrm{V}}$ = $\pi$, depending on the charge ordering of the two particles, while no preferred value exists for electrons originating from other sources. 
The $\psi_{\mathrm{V}}$ distribution for $\mathrm{e}^{+}\mathrm{e}^{-}$ pairs originating from the main dielectron sources, obtained from MC simulations, is shown in Fig.~\ref{fig:PhiVDistribution}-A.

The tagging criterion for the rejection of conversion electrons is defined by $5  < m_{\mathrm{ee}} < 15 \ \mathrm{MeV}/\textit{c}^{2}$, and $0^{\circ} < \psi_{\mathrm{V}}  < 40^{\circ}$ or $140^{\circ} < \psi_{\mathrm{V}}  < 180^{\circ}$.
The correlation between $\psi_{\mathrm{V}}$ and $m_{\mathrm{ee}}$ for electron-positron pairs produced by photon conversion in the material after they have passed the single track rejection is shown in Fig.$~\ref{fig:PhiVDistribution}$-B. The regions used for tagging the conversion candidates are also indicated.  
This background rejection method, in combination with the single-track rejection method, results in a rejection efficiency of conversion electrons from $90\%$ to $80\%$ going from low to high $p_{\mathrm{T}, \mathrm{e}}$, and a maximum rejection probability of signal tracks of $40 \%$ as estimated from simulations, as shown in Fig.$~\ref{fig:ConversionRejectionEfficiency}$.

\begin{figure}[!hbt]
         \centering
                 \includegraphics[height=5.5cm,width=7.5cm]{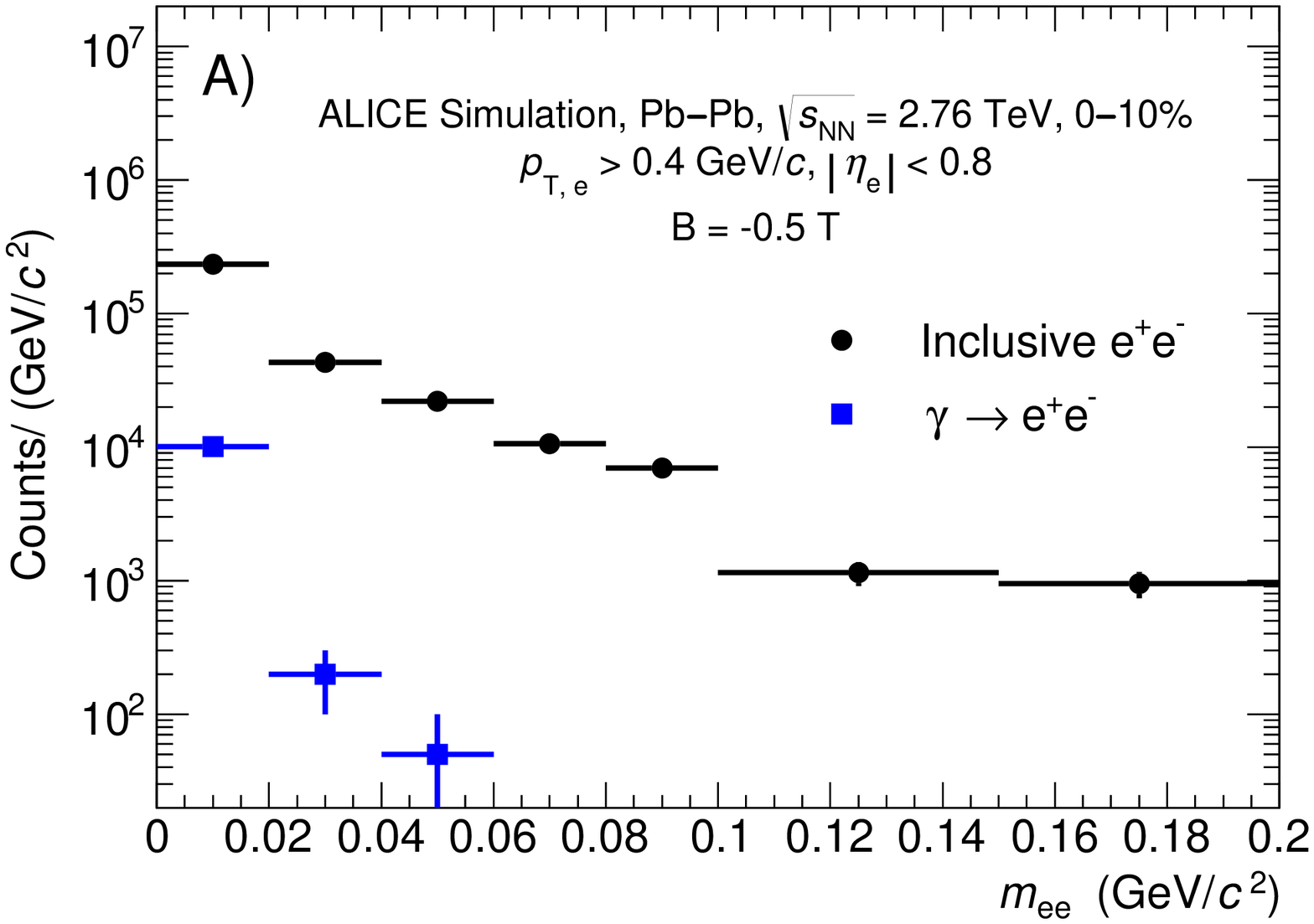}
                 \includegraphics[height=5.5cm,width=7.5cm]{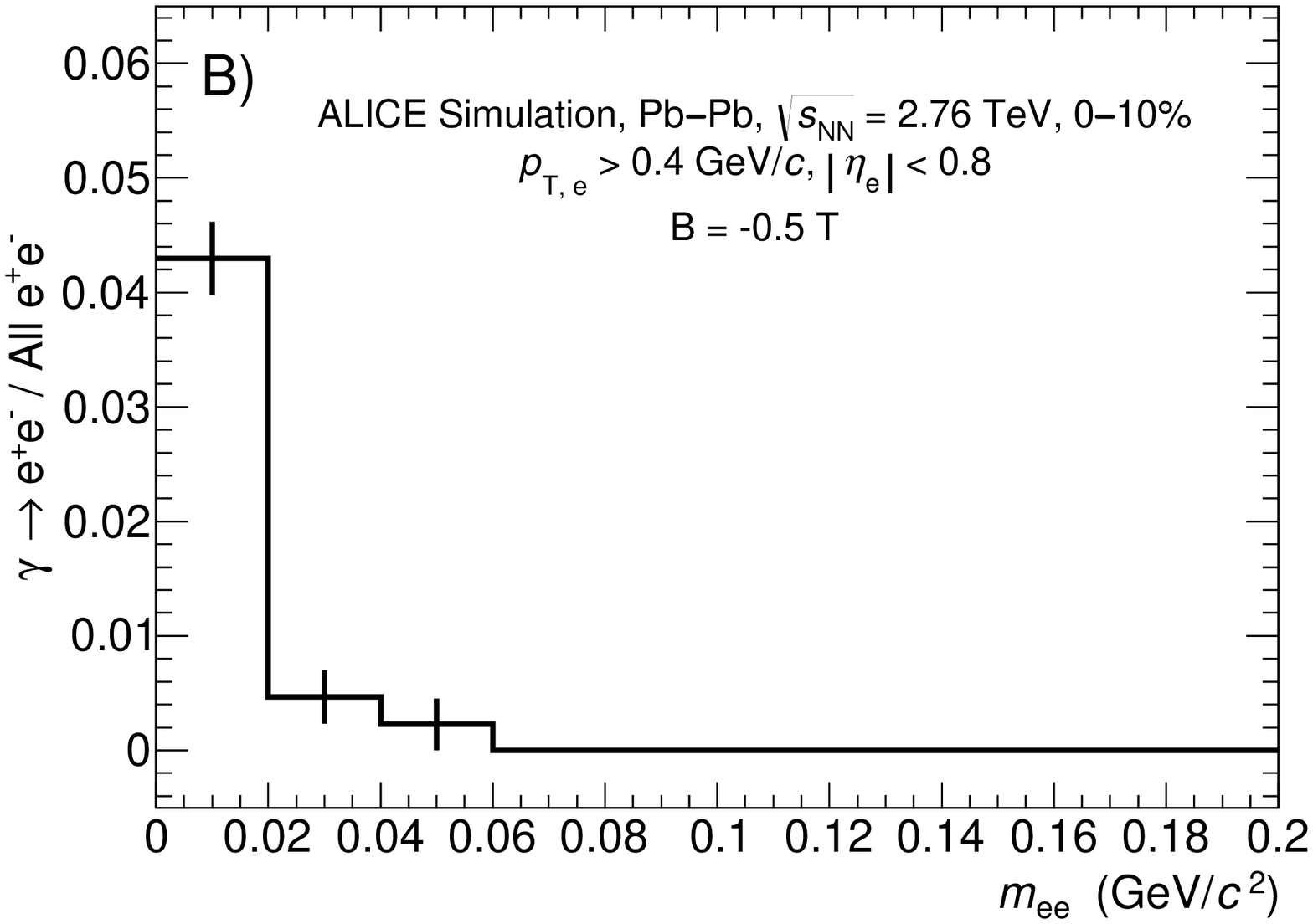}
         \caption{(colour online). (A) Dielectron raw yield, in the low-mass region, in comparison with the residual contribution of conversion electrons, estimated using MC simulations. (B) Relative contribution to the dielectron spectrum of conversion electrons surviving the photon conversion rejection. Only results for the negative magnetic field polarity are shown, with similar values obtained for the positive magnetic field configuration.}
 \label{fig:ResidualConversionContribution}
 \end{figure}

The residual contribution from conversion electrons, estimated from MC simulations, is maximum for $m_{\mathrm{ee}} < 20\ \mathrm{MeV}/\textit{c}^{2}$ and it is negligible (less than $1\%$) at higher masses, as shown in Fig.$~\ref{fig:ResidualConversionContribution}$. The maximum values are smaller than $4 \%$ for the $p_{\mathrm{T}, \mathrm{ee}}$-integrated spectrum, smaller than $3.5 \%$ and $9 \%$  for $1 < p_{\mathrm{T}, \mathrm{ee}} < 2\  \mathrm{GeV}/\textit{c}$ and $2 < p_{\mathrm{T}, \mathrm{ee}} < 4\ \mathrm{GeV}/\textit{c}$, respectively. 
In order to correct for the residual conversion contribution, the dielectron invariant-mass spectrum is scaled by the factor $F = 1 -  (\gamma \rightarrow  \mathrm{e}^{+}\mathrm{e}^{-})/ ({\mathrm{inclusive} \ \mathrm{e}^{+}\mathrm{e}^{-}})$ (see the right panel of Fig.~\ref{fig:ResidualConversionContribution} for the ratio $\gamma \rightarrow  \mathrm{e}^{+}\mathrm{e}^{-}/ {\mathrm{inclusive} \ \mathrm{e}^{+}\mathrm{e}^{-}}$). The results for the negative field polarity are shown for illustration, with similar results for the positive field polarity.

The relative contribution from conversion $\mathrm{e}^{+}\mathrm{e}^{-}$ pairs to the dielectron invariant-mass spectrum is significant up to $m_{\mathrm{ee}} = 0.1\ \mathrm{GeV}/\textit{c}^{2}$ if no conversion rejection is applied. The residual contribution in the mass range $m_{\mathrm{ee}} < 20\ \mathrm{MeV}/\textit{c}^{2}$ is due to conversions occurring in the beam pipe and in the first layer of the ITS, which cannot be removed using the rejection techniques presented above.

\subsection{Background description and signal extraction}
\label{subsec:BackgroundDescriptionAndSignalExtraction}

The combinatorial background is described using the invariant-mass distribution of like-sign dielectron pairs. This is given by the geometric mean of positive and negative like-sign pairs multiplied by two

\begin{equation}
B = 2 \cdot \sqrt{\left[ {\mathrm{d}N}/{\mathrm{d}m_{\mathrm{ee}}} \right]_{++} \cdot \left[ {\mathrm{d}N}/{\mathrm{d}m_{\mathrm{ee}}} \right]_{--}}.
\end{equation}

This distribution describes both the uncorrelated and correlated background. The former arises from uncorrelated electron-positron pairs, while the latter originates from kinematically correlated pairs, such as `cross pairs' from double Dalitz decays and electron-positron pairs produced in the decays of different hadrons inside the same jets or in back-to-back jets.
A small charge asymmetry is observed, manifested in small distortions in the momentum, $\eta$ and $\varphi$ distributions of oppositely-charged particles and a small excess of particles with a given charge. This asymmetry originates from detector geometric inhomogeneities mainly due to the partial installation of the Transition Radiation Detector (TRD), with 13 out of 18 modules in operation, and to some damaged pixels of the SPD which created gaps in its azimuthal coverage.
The TRD, which is placed between the TPC and the TOF detectors, is not used in this analysis. However, it affects the detector acceptance and the TPC-TOF track matching since electrons passing through the TRD material have a larger probability to be absorbed or deflected compared to electrons passing through the gaps. 
These geometrical effects create different acceptances for opposite-sign and like-sign dielectron pairs, requiring an acceptance correction for the latter before it could be used as background estimator. 
The acceptance correction, the so-called $R$-factor, is defined as the ratio between unlike-sign and like-sign dielectrons from mixed events

\begin{equation}
R = \frac{  \left[ {\mathrm{d}N}/{\mathrm{d}m_{\mathrm{ee}}} \right]^{\mathrm{mix}}_{+-} }{   2 \cdot \sqrt { \left[ {\mathrm{d}N}/{\mathrm{d}m_{\mathrm{ee}}} \right]^{\mathrm{mix}}_{++} \cdot   \left[ {\mathrm{d}N}/{\mathrm{d}m_{\mathrm{ee}}}. \right]^{\mathrm{mix}}_{--}} }  
\end{equation}

In the event mixing procedure, events with similar global properties are grouped together sorting them based on the $\textit{z}$-coordinate of the vertex and the centrality of the collision.  
The borders of intervals used to sort the events in the mixing procedure are indicated in Tab.~\ref{tab:EventMixingCells}.

\begin{table}[!hbt]
\centering
\renewcommand{\arraystretch}{1.6}
\begin{tabular}{lclcl}
\hline
     Centrality ($\%$)     &     $\textit{z}$-coordinate of the vertex (cm)     \\
\hline
     [0, 2.5, 5, 7.5, 10]     &     [-10, -6, -2, 2, 6, 10]          \\
\hline
\end{tabular}
\caption{ Borders of intervals used to sort the events in the event-mixing procedure.}
\label{tab:EventMixingCells}
\end{table}

\begin{figure}[!hbt]
         \centering 
 	        \includegraphics[height=5.5cm,width=7.5cm]{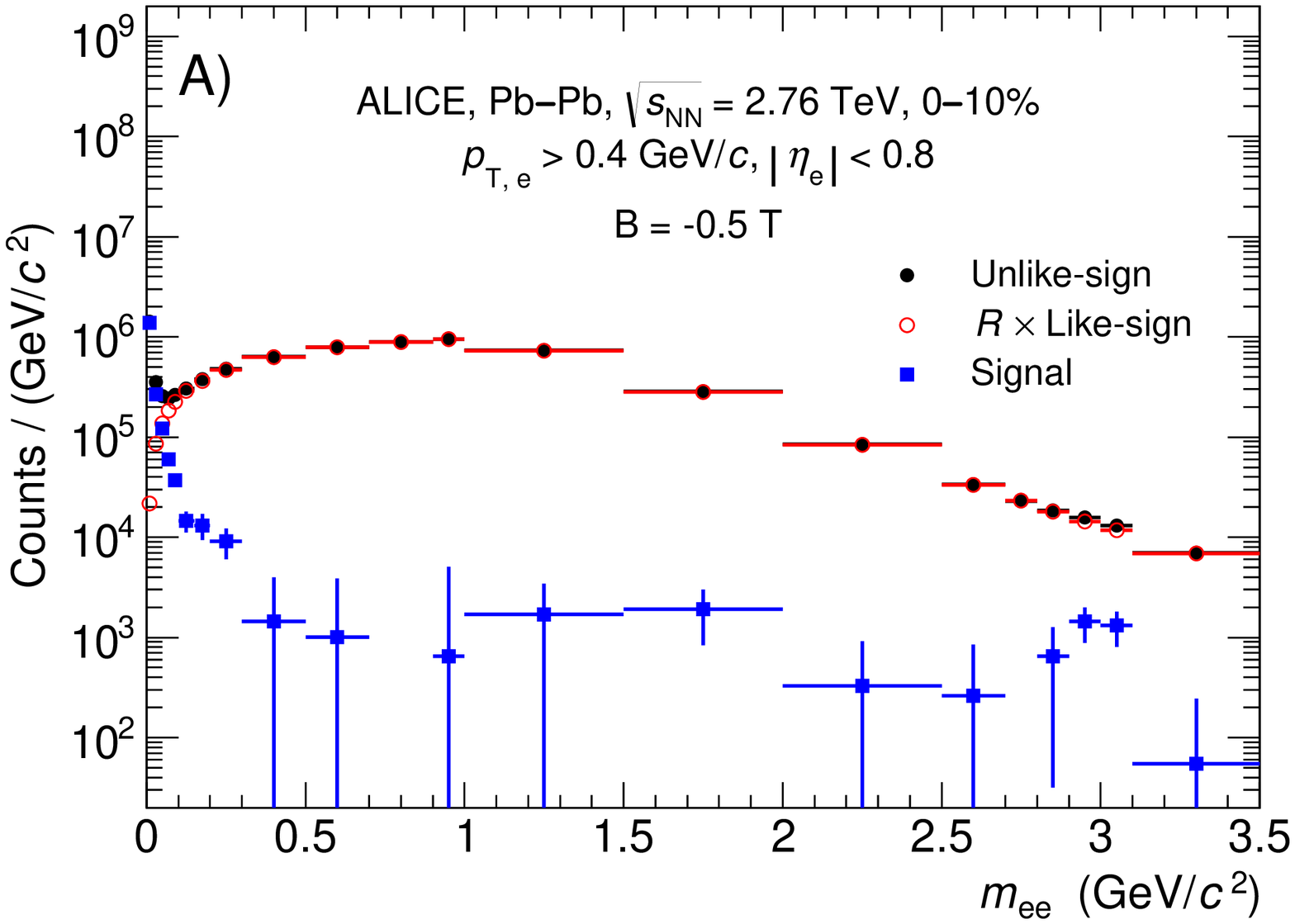}          
                 \includegraphics[height=5.5cm,width=7.5cm]{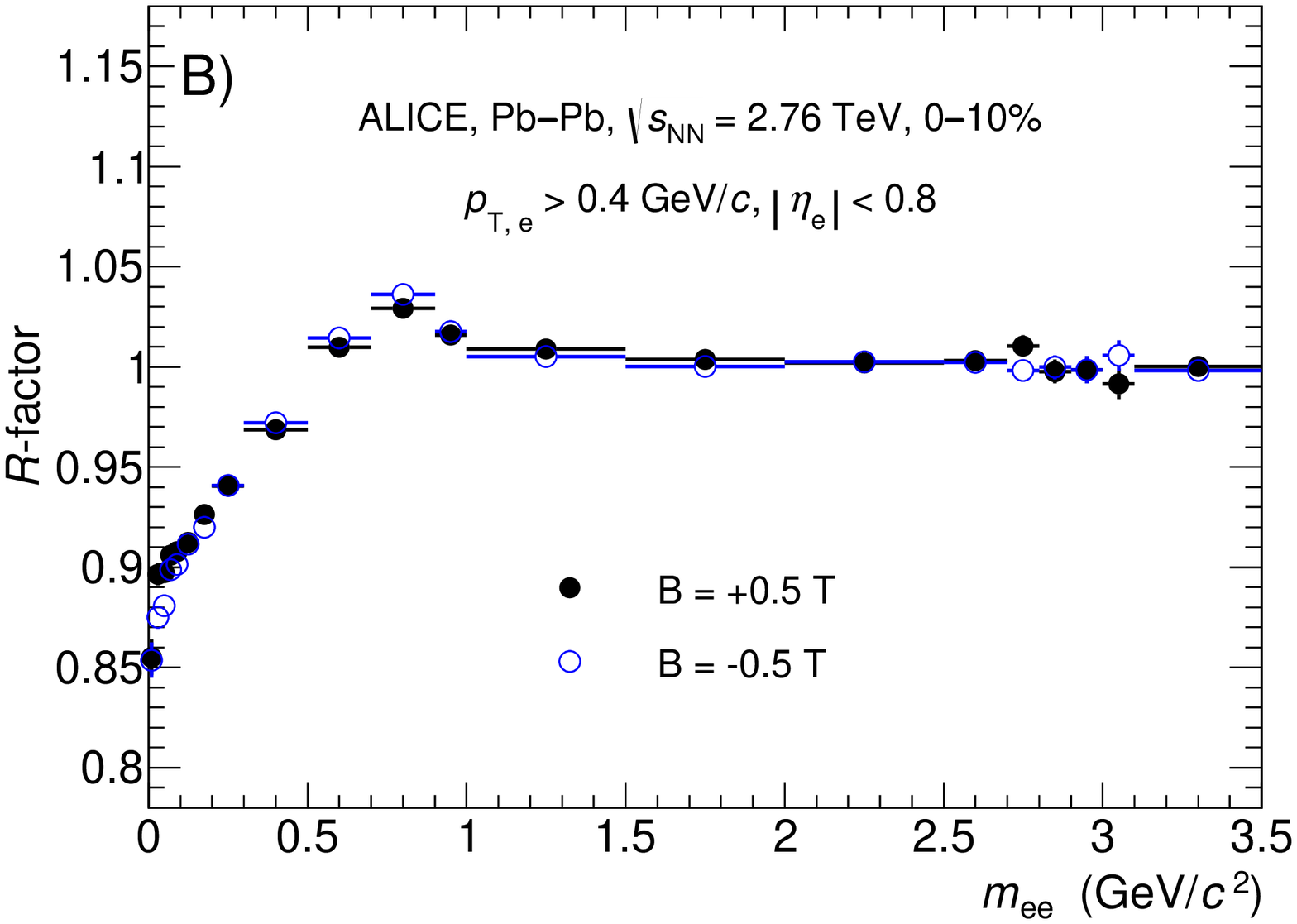}
         \caption{(colour online). (A) Invariant-mass distributions of unlike-sign (black), like-sign (red) and signal (blue) dielectron pairs. (B) Acceptance correction factor ($R$-factor) for both magnetic field orientations. Only statistical uncertainties are shown. }
 \label{fig:combinatorialBackground}
 \end{figure}

The invariant-mass spectrum of signal dielectron pairs is obtained by subtracting the combinatorial background estimated from the $R$ $\times$ like-sign spectrum from the unlike-sign spectrum. 
The invariant-mass spectra of unlike-sign, $R$ $\times$ like-sign and signal dielectron pairs are shown in Fig.~\ref{fig:combinatorialBackground}-A, while the $R$-factor is shown in Fig.~\ref{fig:combinatorialBackground}-B. 
The effects of charge asymmetry on oppositely-charged particles are reversed for different magnetic field orientations. This is reflected in the different shapes of the acceptance correction factor which are smaller than about $2\%$.

\begin{figure}[!hbt]
         \centering 
 	        \includegraphics[height=8cm,width=11cm]{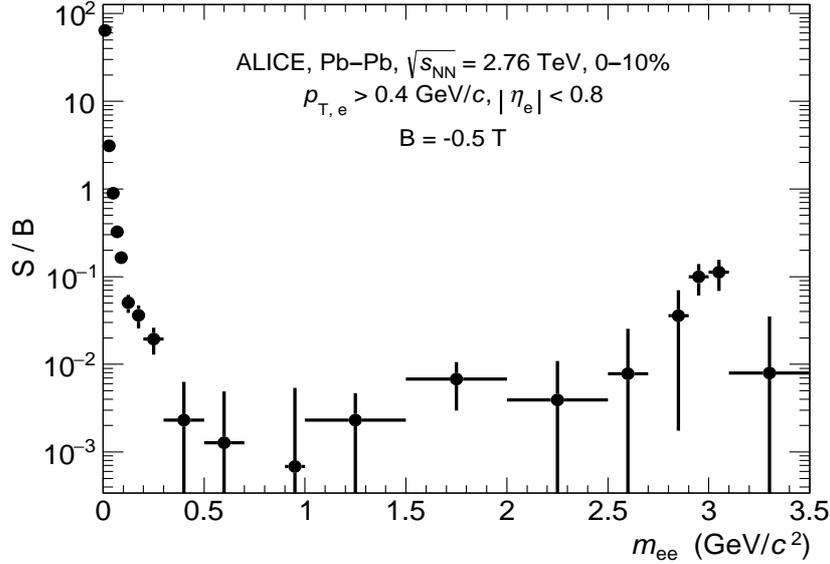}          
         \caption{Signal-to-background ratio of the dielectron measurement as a function of the invariant mass for the negative magnetic field polarity. Similar values are obtained for the positive magnetic field configuration. }
 \label{fig:SignalToBackground}
 \end{figure}

The signal-to-background ratio of the dielectron spectrum as a function of the invariant mass, shown in Fig.~\ref{fig:SignalToBackground} for the $p_{\mathrm{T}, \mathrm{ee}}$-integrated spectrum, is a factor approximately 2 larger compared to the case where no conversion rejection is applied.

\newpage
\subsection{Pair reconstruction efficiency}
\label{subsec:PairReconstructionEfficiency}

\begin{figure}[!hbt]
         \centering
                 \includegraphics[height=8cm,width=10cm]{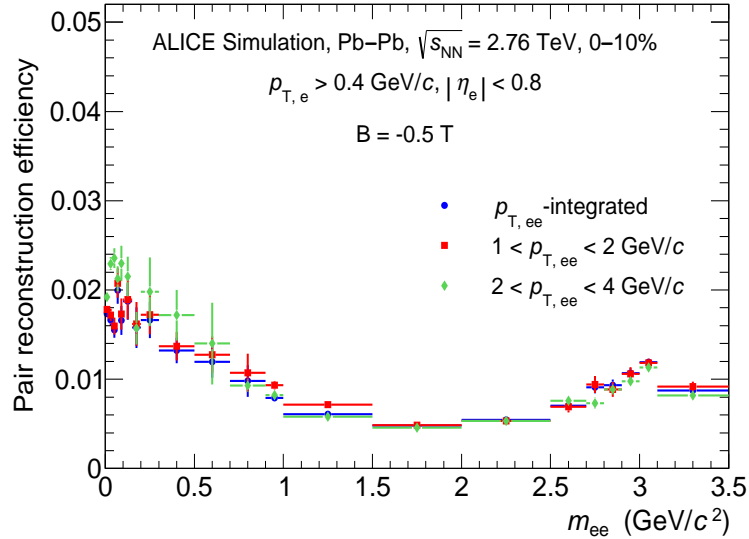}
         \caption{(colour online). Reconstruction efficiency of dielectrons as a function of the invariant mass for different $p_{\mathrm{T}, \mathrm{ee}}$ intervals. The statistical uncertainties are represented by the vertical bars. Only results for the negative magnetic field polarity are shown, with similar values obtained for the positive magnetic field configuration. }
\label{fig:PairEfficiency}
 \end{figure}

The pair reconstruction efficiency is calculated as the ratio between the invariant-mass distributions of reconstructed and generated dielectron signal pairs in the simulation.
The reconstructed momentum of electrons is affected by the finite momentum resolution of the detector, the tracking procedure, and the energy loss by bremsstrahlung in the detector material. 
These effects are described in the simulation of the interactions between generated particles and the experimental apparatus, modeled by GEANT 3 \cite{GEANT}. 
The detector effects are applied to transform the generated momentum of electrons $\vec{p}_{\mathrm{gen}, \mathrm{e}}$ into the corresponding `measurable' momentum $\vec{p}_{\mathrm{rec}, \mathrm{e}}$, i.e. the momentum that would be measured by the detector. The latter is used to calculate the invariant mass of generated dielectron pairs in the denominator of the efficiency. 
\newline
The procedure used in the electron's momentum transformation, which is outlined in detail in \cite{MomentumTransformationMatrix}, is based on the detector response matrices obtained from simulations which connect the components of the momentum vector of generated and reconstructed electrons. The detector behavior is fully described by four transformation matrices, one for $p_{\mathrm{T}, \mathrm{e}}$, one for the polar angle ($\theta_{\mathrm{e}}$) and two for the azimuthal angle ($\varphi_{\mathrm{e}^{+}}$ and $\varphi_{\mathrm{e}^{-}}$) of electrons and positrons. 
\newline 
The MC simulations used to calculate the pair reconstruction efficiency include dielectron sources produced by HIJING and an additional sample of dielectron sources injected into the simulated events. This injected sample includes $\pi^{0}$, $\eta$, $\eta^{\prime}$, $\rho^{0}$, $\omega$, $\phi$ and J/$\psi$, forced to decay into dielectrons, produced in equal amounts with uniform $p_{\mathrm{T}}$ distributions, and an enriched sample of heavy-flavor sources with forced semileptonic decay channel.
The input $p_{\mathrm{T}}$ distributions of dielectron sources are corrected using $p_{\mathrm{T}}$-dependent weights, which include a correction factor to adjust their relative particle abundances according to measurements \cite{PizeroMeasurementPbPb,KToPiRatio,mtScaling}.
The $p_{\mathrm{T}}$-differential cross section of neutral pions measured by ALICE \cite{PizeroMeasurementPbPb} is used as a reference to modify the input spectrum of $\pi^{0}$, while the shapes of $p_{\mathrm{T}}$ distributions of all other light-flavor mesons are obtained from $m_{\mathrm{T}}$-scaling of the $\pi^{0}$ spectrum, replacing $p_{\mathrm{T}}$ with $\sqrt{m^{2}-m_{\pi}^{2} + \left(    p_{\mathrm{T}}/\textit{c} \right)^{2}}$. 
For the enriched heavy-flavor sources, the input distributions are tuned using PYTHIA 6  \cite{PYTHIA} (Perugia 2011 tune).
The correlated semileptonic heavy-flavor decays produced by HIJING are used with unmodified input spectra. The latter choice is justified by the consistency between the shape of the invariant-mass distribution of dielectrons from heavy-flavor decays generated by HIJING and PYTHIA 6. 
The pair reconstruction efficiency includes a correction factor to account for the different random rejection probability, between data and MC simulations, of signal tracks in the photon conversion rejection method based on pair correlations (see Sec.~\ref{subsec:ConversionRejectionMethods}). This correction factor is given by the ratio between the rejection probability of dielectron signal pairs embedded into real and simulated events. 
The reconstruction efficiency of dielectron signal pairs as a function of their invariant mass for different $p_{\mathrm{T}, \mathrm{ee}}$ intervals is shown in Fig.$~\ref{fig:PairEfficiency}$.

\subsection{Hadronic cocktail}
\label{subsec:HadronicCocktail}

The hadronic cocktail, obtained from simulations, contains the sum of the contributions to the dielectron spectrum originating from hadron decays. Similarly to the efficiency calculation, the $p_{\mathrm{T}}$-spectrum of $\pi^{0}$ measured in Pb--Pb collisions at $\mathbf{\sqrt{{\textit{s}}_{\mathrm{NN}}}}$ = 2.76 TeV \cite{PizeroMeasurementPbPb} is used as input parametrization to the MC simulation to generate the invariant-mass spectrum of dielectrons from $\pi^{0}$-Dalitz decays, while $m_{\mathrm{T}}$-scaling is applied to generate the contributions of the remaining light-flavored mesons, except for the $\eta$. The contribution from the latter is estimated as the average of the spectra obtained using the parametrizations retrieved from the $\eta/\pi^{0}$ ratio as a function of $p_{\mathrm{T}}$ measured in pp collisions at $\sqrt{\textit{s}}=$ 7 TeV \cite{EtaToPiRatio} and from the $K^{\pm} / \pi^{\pm}$ ratio measured in Pb--Pb collisions at $\mathbf{\sqrt{{\textit{s}}_{\mathrm{NN}}}}$ = 2.76 TeV \cite{KToPiRatio}, respectively. In the latter, kaons are used as proxy for the $\eta$, due to their similar masses. 
This parametrization of the $\eta$ spectrum is in agreement with the recently published $p_{\mathrm{T}}$ spectrum measured by ALICE in Pb--Pb collisions at $\mathbf{\sqrt{{\textit{s}}_{\mathrm{NN}}}}$ = 2.76 TeV \cite{EtaToPiRatioPbPbALICE}.
\newline
The contributions from semileptonic charm and beauty decays are obtained from PYTHIA 6 (Perugia 2011 tune) for pp collisions at $\sqrt{\textit{s}}$ = 2.76 TeV and scaled by the average number of binary collisions ($\langle N_{\mathrm{coll}} \rangle$) obtained from the MC Glauber model \cite{GlauberModel}. 
The invariant-mass distributions of dielectrons from charm and beauty decays obtained using PYTHIA 6 are normalized using the cross sections $\sigma_{\mathrm{c\overline{c}}}$ and $\sigma_{\mathrm{b\overline{b}}}$ measured in pp collisions at  $\mathbf{\sqrt{{\textit{s}}}}$ = 2.76 TeV \cite{charmMeasurement, beautyMeasurement}. 
The average branching ratios of $9.6\%$ for the $\mathrm{c} \rightarrow \mathrm{e}$ decays \cite{BRcharm} and of $21.5\%$ for the $\mathrm{b} (\rightarrow \mathrm{c}) \rightarrow \mathrm{e}$ decays \cite{PDG} are used in the PYTHIA 6 simulation. 
This approach, based on the $N_{\mathrm{coll}}$ scaling, assumes no shadowing and no energy loss of charm and beauty quarks in the medium. 
An alternative method to obtain the charm contribution is also adopted, which is based on the randomization of the initial angular correlations of $\mathrm{c\overline{c}}$ pairs to simulate their interactions with the medium \cite{dielectronsAuAuPHENIX}. More specifically, electrons and positrons are generated independently starting from the input $p_{\mathrm{T}, \mathrm{e}}$, $\eta_{\mathrm{e}}$ and $\varphi_{\mathrm{e}}$ distributions of single electrons from charm decays produced by PYTHIA 6 (Perugia 2011 tune), thus ignoring their initial correlations in the azimuthal angle and pseudorapidity. The comparison between data and the two cocktail versions using different approaches to estimate the charm contribution is illustrated in Fig.~\ref{fig:dielectronSpectrumRandomizedCharm} below.
\newline
The J/$\psi$ contribution is generated in pp collisions and then scaled by $\langle N_{\mathrm{coll}} \rangle$ and by the $R_{AA}$ measured in central (0$-$10$\%$) Pb--Pb collisions at $\mathbf{\sqrt{{\textit{s}}_{\mathrm{NN}}}}$ = 2.76 TeV by ALICE \cite{RAAJPsi}.

\subsection{Systematic uncertainties}
\label{subsec:SystematicUncertainties}

The systematic uncertainties on the measured dielectron spectrum originate from tracking, particle identification, photon conversion rejection, hadron contamination and background subtraction. 
The former three contributions are mainly due to the non-perfect description of the detector and of real data by MC simulations that are used to calculate the pair reconstruction efficiency. The systematic uncertainty on hadron contamination is connected to the uncertainty in the purity estimation and to the uncertainty on the relative contribution of correlated electron-hadron and hadron-hadron pairs, where hadrons are wrongly identified as electrons (see Sec.~\ref{subsec:ElectronIdentificationAndHadronContamination}). The systematic uncertainty on the background subtraction is related to the precision in the estimation of the combinatorial background.
To estimate the magnitude of systematic uncertainties, the dielectron spectrum is studied by varying the parameters of the analysis used for the track selection, particle identification, and photon conversion rejection. 
Using different selection criteria, a variation of a factor larger than 10 in the pair reconstruction efficiency and of a factor larger than 2 in the S/B are obtained, respectively, such that also the uncertainties from the background subtraction are included. 
The dielectron spectra obtained for the different settings are corrected for the corresponding pair reconstruction efficiencies, obtained using the same settings in the simulation. 
The systematic uncertainties are calculated as the RMS of the spread of efficiency-corrected dielectron yields in which all contributions are considered together to take into account their correlations. 
In order to obtain better statistical precision and to minimize the interference between statistical and systematic uncertainties, the data points are grouped into three mass ranges:  $0 < m_{\mathrm{ee}} < 0.5\ \mathrm{GeV}/\textit{c}^{2}$, $0.5 < m_{\mathrm{ee}} < 2.8\ \mathrm{GeV}/\textit{c}^{2}$ and $2.8 < m_{\mathrm{ee}} < 3.5\ \mathrm{GeV}/\textit{c}^{2}$. The systematic uncertainties are found to be weakly dependent on $p_{\mathrm{T}, \mathrm{ee}}$ and are estimated using the $p_{\mathrm{T}, \mathrm{ee}}$-integrated spectrum.
The relative systematic uncertainties are calculated for each mass region and they are assigned to the data points in the corresponding mass ranges. 
The systematic uncertainty on the relative contribution of hadron contamination is assumed to be of the order of $50\%$, which results in a contribution of less than $1-2\%$ on the dielectron spectra which is added in quadrature to the other contributions.
Table ~\ref{tab:SystematicUncertainties} summarizes the systematic uncertainties in the different mass regions.

\begin{table}[!hbt]
\centering
\renewcommand{\arraystretch}{1.6}
  \begin{tabular}{lcP{5cm}lcP{5cm}l}
\hline
     Mass range (GeV/$\textit{c}^{2}$)    & Rel. syst. uncert. ($\%$)   \\       
\hline
     [0, 0.5]             &     7      \\         \hline
     [0.5, 2.8]       &    35     \\        \hline
      [2.8, 3.5]    &    15   \\       \hline
\end{tabular}
\caption{Relative systematic uncertainties in different mass ranges. }
\label{tab:SystematicUncertainties}
\end{table}

Regarding the systematic uncertainties on the hadronic cocktail, the upper and lower limits of the invariant-mass distributions of the light-flavor component are obtained by simulating the hadron decays using as input the parametrizations of neutral pions corresponding to the upper and lower limits of their systematic uncertainties. 
The systematic uncertainty on the heavy-flavor component of the hadronic cocktail is obtained by adding in quadrature the contributions on the cross sections $\mathrm{c}\overline{\mathrm{c}}$ and $\mathrm{b}\overline{\mathrm{b}}$, 
that on the total inelastic cross section in pp collisions at $\sqrt{\textit{s}}=2.76\ \mathrm{TeV}$ \cite{sigmaInelasticTotalpp}, that on $\langle N_{\mathrm{coll}} \rangle$ and on the branching ratios of charm decays. The latter two contributions are $24\%$ and $7\%$, respectively \cite{BRcharm,PDG}. The systematic uncertainty on the J/$\psi$ $R_{AA}$ is also included.

\section{Results}
\label{sec:Results}

\subsection{Cocktail comparison}
\label{subsec:CocktailComparison}

\begin{figure}[!hbt]
         \centering 
 	        \includegraphics[height=9cm,width=11cm]{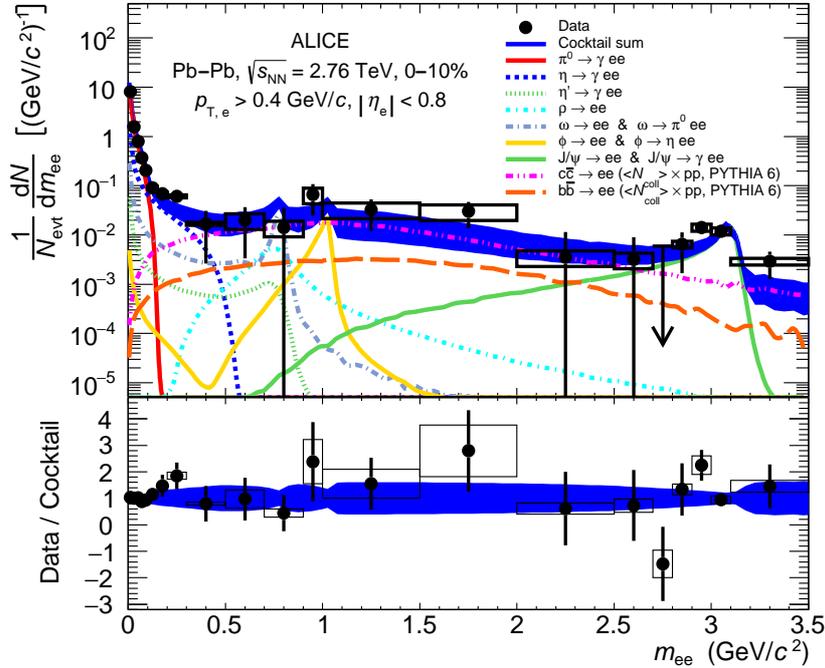}          
         \caption{(colour online). Dielectron invariant-mass spectrum measured in central (0$-$10$\%$) Pb--Pb collisions at $\mathbf{\sqrt{{\textit{s}}_{\mathrm{NN}}}}$ = 2.76 TeV in comparison with the hadronic cocktail. The statistical and systematic uncertainties of the data are represented by vertical bars and boxes. The arrow for one mass range represents the upper limit at $90\%$ C.L. The blue band represents the systematic uncertainties on the hadronic cocktail. }
 \label{fig:dielectronSpectrum}
 \end{figure}

The measurement of the dielectron invariant-mass spectrum is shown in Fig.~\ref{fig:dielectronSpectrum} in comparison with the expected contributions from hadronic sources (hadronic cocktail). In the mass range $2.7 < m_{\mathrm{ee}}<2.8\ \mathrm{GeV/\textit{c}^{2}}$, where a negative dielectron yield is measured, an upper limit at $90\%$ C.L. is set using the Feldman and Cousins methodology \cite{FeldmanCousins}, considering the statistical and systematic uncertainties as uncorrelated. In the data-to-cocktail ratio, the statistical uncertainties on the dielectron spectrum and on the hadronic cocktail are added in quadrature, while systematic uncertainties are shown separately. In particular, the systematic uncertainty on the hadronic cocktail is represented by the blue band along the cocktail line shape.

The contribution to the dielectron spectrum from the $\rho^{0}$ meson is expected to be modified due to its significant regeneration during the hadronic phase and the broadening of its mass shape. 
The ratio of data and the hadronic cocktail, excluding the contribution from the vacuum $\rho^{0}$, is calculated in the invariant-mass range $0.15 < m_{\mathrm{ee}}<0.7\ \mathrm{GeV}/\textit{c}^{2}$, where an excess of the dielectron yield is observed with respect to the cocktail of hadronic sources in other dilepton experiments, and its value is $R = 1.40 \pm 0.28\ (\mathrm{stat}) \pm 0.08\ (\mathrm{syst}) \pm 0.27\ (\mathrm{cocktail})$. 
This ratio corresponds to a larger dielectron production with respect to the contribution from hadronic sources excluding the vacuum $\rho^{0}$ with a statistical significance of $(R-1)/ \Delta R = 1.41$, being $\Delta R$ the statistical uncertainty on the data-to-cocktail ratio. The limited sensitivity of the dielectron measurement in the low-mass region, due to the low number of events and to the small signal-to-background ratio in the invariant-mass range $0.15 < m_{\mathrm{ee}}<0.7\ \mathrm{GeV}/\textit{c}^{2}$, prevents any analysis of a possible excess spectrum. 
We consider in the following in more detail the compatibility of the data with thermal radiation emission at the level that is expected from the theoretical model that described successfully the SPS and RHIC data \cite{signaturesThermalDileptonsRapp,dileptonSpectroscopyRapp}.

\subsection{Thermal dielectrons}
\label{subsec:ThermalDielectrons}

\begin{figure}[!hbt]
         \centering
                 \includegraphics[height=9cm,width=11cm]{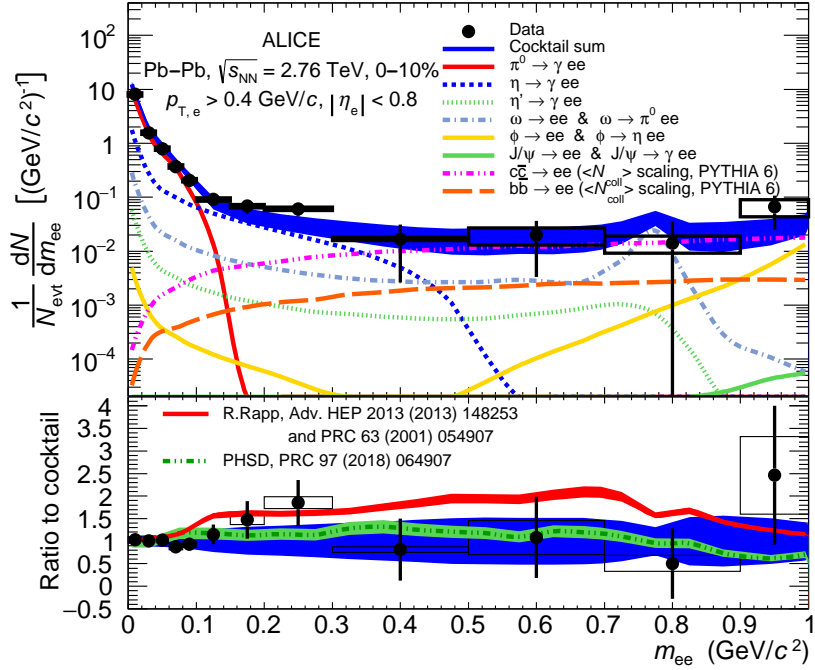}
   \caption{(colour online). Dielectron invariant-mass spectrum for $m_{\mathrm{ee}}<1\ \mathrm{GeV/\textit{c}^{2}}$ in comparison with the predictions from R. Rapp \cite{signaturesThermalDileptonsRapp,dileptonSpectroscopyRapp} and PHSD \cite{PHSD} theoretical models. The statistical and systematic uncertainties of data are represented by vertical bars and boxes. The systematic uncertainties on the theory curves are due to the energy loss correction that is applied based on a cocktail simulation (see text for details).}
\label{fig:TheoryComparison}
\end{figure}

The dielectron invariant-mass spectrum for $m_{\mathrm{ee}}<1\ \mathrm{GeV/\textit{c}^{2}}$ is compared to the expectations from two theoretical model calculations that include thermal dielectrons from the partonic and hadronic phases. 
Both models predict a broadening of the $\rho^{0}$ electromagnetic spectral function as an effect of interactions in the hot hadron gas phase. This effect is directly connected to the partial restoration of chiral symmetry at high temperatures close to the phase boundary. 
In the first model, from R. Rapp  \cite{signaturesThermalDileptonsRapp,dileptonSpectroscopyRapp}, the thermal component is obtained from an expanding fireball model for central Pb--Pb collisions at $\mathbf{\sqrt{{\textit{s}}_{\mathrm{NN}}}}$ = 2.76 TeV, corresponding to $\langle \mathrm{d}N_{\mathrm{ch}}/\mathrm{d}y \rangle=1600$, using a lattice-QCD inspired approach with an equation of state for the QGP with critical temperature $T_{\mathrm{c}}=170\ \mathrm{MeV}$. 
The thermal emission rate of dielectrons from the hadronic phase is calculated based on the hadronic many-body theory with in-medium modified $\rho^{0}$ \cite{signaturesThermalDileptonsRapp,dileptonSpectroscopyRapp}. The same approach is used to describe RHIC and SPS data.  
\newline
The second model from T. Song $\textit{et al.}$ \cite{PHSD} is based on the Parton-Hadron-String Dynamics (PHSD) transport approach. The contribution of thermal dielectrons from the partonic phase is calculated assuming that the degrees of freedom in the QGP are massive off-shell strongly interacting quasi-particles. In the hadronic phase, the dielectron production is calculated using in-medium modified electromagnetic spectral functions of low-mass vector mesons which change dynamically during
the propagation through the medium and evolve towards on-shell spectral functions in the vacuum. 
Dielectrons from the decays of heavy quarks are also included, whose interactions with the medium are described using the Dynamical Quasi-Particle Model (DQPM) \cite{DQPMpartonTransport,DQPMeffectiveInteractions}. 
\newline
The theoretical predictions are obtained using the `true' momentum of electrons, e.g. the momentum that is generated in the simulation, and the electrons are selected in the same kinematic region as in data $0.4 < p_{\mathrm{T, e}}< 5$ GeV/$\textit{c}$ and $|\eta_{\mathrm{e}}|<0.8$. 
\newline
The measured dielectron spectrum is affected by the energy loss of electrons by bremsstrahlung in the interactions with the detector material. These effects produce a shift of the $p_{\mathrm{T, e}}$ spectrum of electrons towards lower values. In real data, the number of electrons with a $p_{\mathrm{T, e}}$ larger than the lower threshold is smaller compared to the ideal case in which electrons are not affected by energy loss. 
For a fair comparison with the data, a correction factor is applied to the theoretical predictions that would otherwise overestimate the data. The correction factor is calculated using a cocktail simulation. This is defined as the ratio between the invariant-mass spectra of dielectrons obtained using the measured and generated momenta. The kinematic range selection is applied on the corresponding measured and generated momenta, respectively.
The correction factor is 0.9,  approximately constant as a function of the invariant mass. A systematic uncertainty of 0.1 is assumed on this correction factor, i.e. 100$\%$ of the difference between the corrected theoretical calculation and that obtained using the generated momentum of electrons, which results in the uncertainty bands represented along the theory curves.
The dielectron spectrum in the invariant-mass range $m_{\mathrm{ee}}<1\ \mathrm{GeV/\textit{c}^{2}}$ is shown in comparison with the predictions from these two models in Fig.$~\ref{fig:TheoryComparison}$.

The contribution from the hadronic cocktail is slightly underestimated in the PHSD model for $m_{\mathrm{ee}}>0.8\ \mathrm{GeV/\textit{c}^{2}}$, which results in a ratio to our hadronic cocktail which is smaller than 1. 
In the invariant-mass range of interest ($m_{\mathrm{ee}}<1\ \mathrm{GeV/\textit{c}^{2}}$), data are consistent with the predictions from both theoretical models within the experimental uncertainties.

\newpage
\subsection{Charm contribution}
\label{subsec:CharmContribution}

\begin{figure}[!hbt]
         \centering 
 	        \includegraphics[height=9cm,width=11cm]{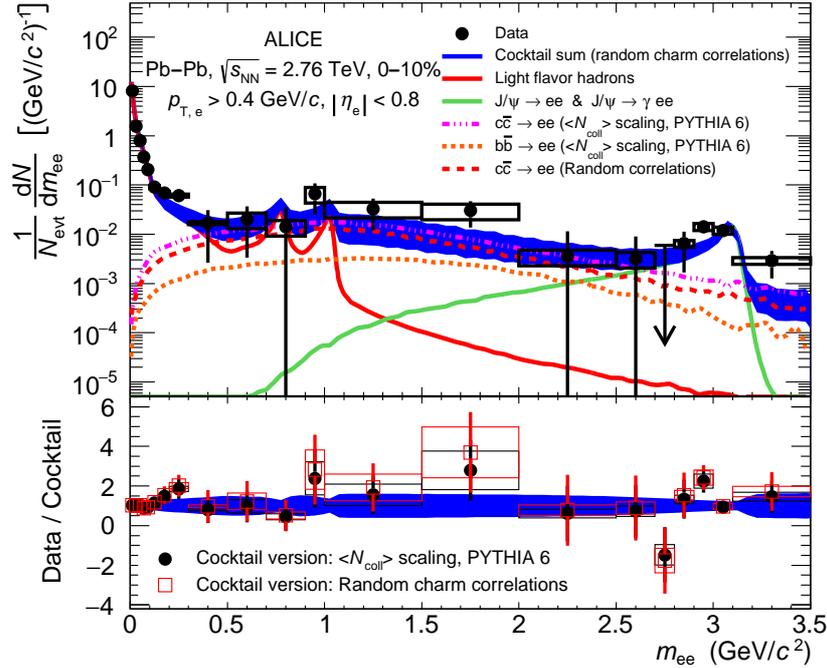}                      
         \caption{(colour online). Dielectron invariant-mass spectrum compared to the two different versions of the hadronic cocktail. In one version, shown with a dash-dotted magenta line, the charm contribution is obtained from PYTHIA via $\textit{N}_{\mathrm{coll}}$-scaling, while in the other version, shown with a dotted red line, a random correlation of dielectrons from charm decays is assumed. The statistical and systematic uncertainties of data are represented by vertical bars and boxes.}
 \label{fig:dielectronSpectrumRandomizedCharm}
 \end{figure}

The contribution to the dielectron invariant-mass spectrum from semileptonic heavy-flavor hadron decays is obtained by assuming that the production cross sections of $\mathrm{c\overline{c}}$ and $\mathrm{b\overline{b}}$ pairs in Pb--Pb collisions at $\mathbf{\sqrt{{\textit{s}}_{\mathrm{NN}}}}$ = 2.76 TeV are given by those in pp collisions at the same center-of-mass energy, scaled by the average number of binary nucleon-nucleon collisions $\langle\textit{N}_{\mathrm{coll}}\rangle$.
This approach ignores shadowing and suppression effects due to the interactions of charm and beauty quarks with other partons in the medium. 
The latter effects are studied for the charm contribution only, which is the dominant source of dielectrons in the intermediate mass region, by assuming that the initial angular correlation of $\mathrm{c\overline{c}}$ pairs is destroyed due to the interactions with the medium. More specifically, electrons and positrons from charm decays are generated independently using the input $p_{\mathrm{T}, \mathrm{e}}$, $\eta_{\mathrm{e}}$ and $\varphi_{\mathrm{e}}$ distributions provided by PYTHIA, thus ignoring the initial correlations imposed by the decay kinematics. 
\newline
In Fig.~\ref{fig:dielectronSpectrumRandomizedCharm} the dielectron spectrum is shown in comparison with two versions of the hadronic cocktail, one containing the charm contribution obtained from PYTHIA via $\textit{N}_{\mathrm{coll}}$-scaling, and the other obtained considering random correlation of dielectrons from charm decays. 
This procedure of angular decorrelation results in a suppression of dielectrons in the intermediate mass range approximately by a factor of 2 compared to the $\textit{N}_{\mathrm{coll}}$-scaling scenario due to a larger number of electrons generated outside of the acceptance, i.e. electrons with $|\eta_{\mathrm{e}}|>0.8$.
\newline
The limited statistical precision of the dielectron measurement in the mass region dominated by heavy-flavor decays and the systematic uncertainties on the $\mathrm{c\overline{c}}$ cross section prevent any conclusion on the effects of interactions between heavy quarks and other partons in the medium as both versions of the hadronic cocktail are consistent with the data within the uncertainties.

\newpage
\subsection{Virtual direct photons}
\label{subsec:VirtualDirectPhotons}

The fraction of virtual direct photons over inclusive photons is measured in the kinematic region $p_{\mathrm{T, ee}} \gg m_{{\mathrm{ee}}}$ (quasi-real virtual photons) by fitting the data using $\chi^{2}$ minimization in the invariant-mass range $100<m_{\mathrm{ee}}<300\ \mathrm{MeV}/\textit{c}^{2}$, for the transverse-momentum intervals $1<p_{\mathrm{T}, \mathrm{ee}}<2\ \mathrm{GeV}/\textit{c}$ and $2<p_{\mathrm{T}, \mathrm{ee}}<4\ \mathrm{GeV}/\textit{c}$, using a three-component function 

\begin{equation}
f(m_{\mathrm{ee}}) = r\cdot f_{\mathrm{dir}} (m_{\mathrm{ee}}) + (1-r) \cdot f_{\mathrm{LF}} (m_{\mathrm{ee}}) + f_{\mathrm{HF}} (m_{\mathrm{ee}}).
\label{eq:ThreeComponentFunc}
\end{equation}

In the above equation, $f_{\mathrm{dir}} (m_{\mathrm{ee}})$ is the expected invariant-mass distribution of virtual direct photons, described by the Kroll-Wada equation \cite{KrollWada}, $f_{\mathrm{LF}} (m_{\mathrm{ee}})$ and $f_{\mathrm{HF}} (m_{\mathrm{ee}})$ are the mass distributions of the light-flavor and heavy-flavor components of the hadronic cocktail, respectively. Each component in Eq.~\ref{eq:ThreeComponentFunc} is integrated over the bin width in each invariant-mass range. The charm contribution obtained from PYTHIA 6 via $N_{\mathrm{coll}}$-scaling is used for the fit. 
The spectra of $f_{\mathrm{LF}} (m_{\mathrm{ee}})$ and $f_{\mathrm{dir}} (m_{\mathrm{ee}})$ are independently normalized to data in the mass range $0<m_{\mathrm{ee}}<20\ \mathrm{MeV}/\textit{c}^{2}$ before the fit is executed. 
The fit parameter $\textit{r}$, which was varied in the range [0,1] in the $\chi^{2}$ minimization procedure, represents the fraction of virtual direct photons     

\begin{equation}
r = \left[ \frac{\gamma^{*}_{\mathrm{dir}} }{\gamma^{*}_{\mathrm{inclusive}}} \right]_{ (m_{\mathrm{ee}} \rightarrow 0\ \mathrm{GeV}/\textit{c}^{2})}. 
\end{equation}

\begin{figure}[!hbt]
         \centering
                 \includegraphics[height=5.5cm,width=7.5cm]{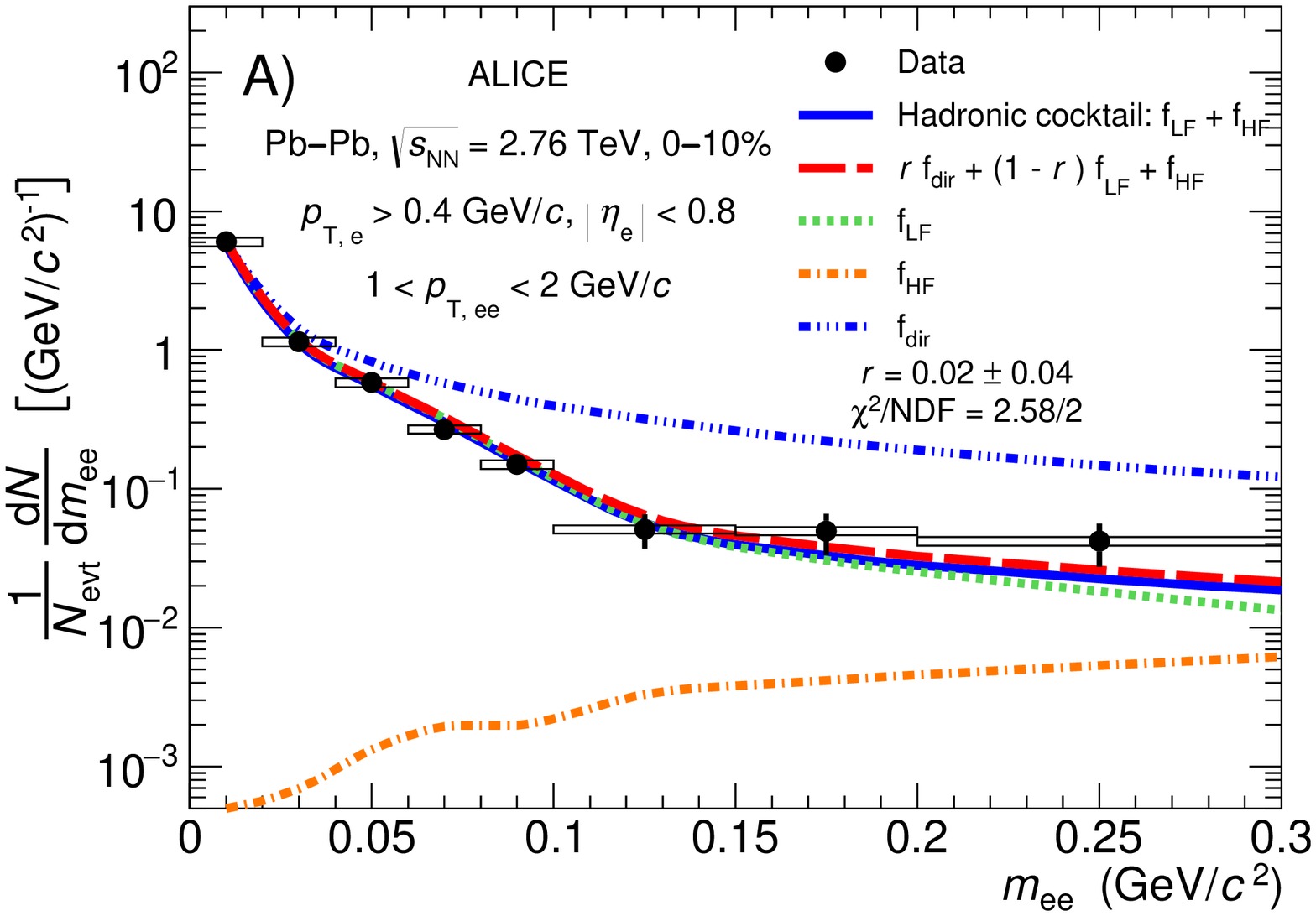}
                 \includegraphics[height=5.5cm,width=7.5cm]{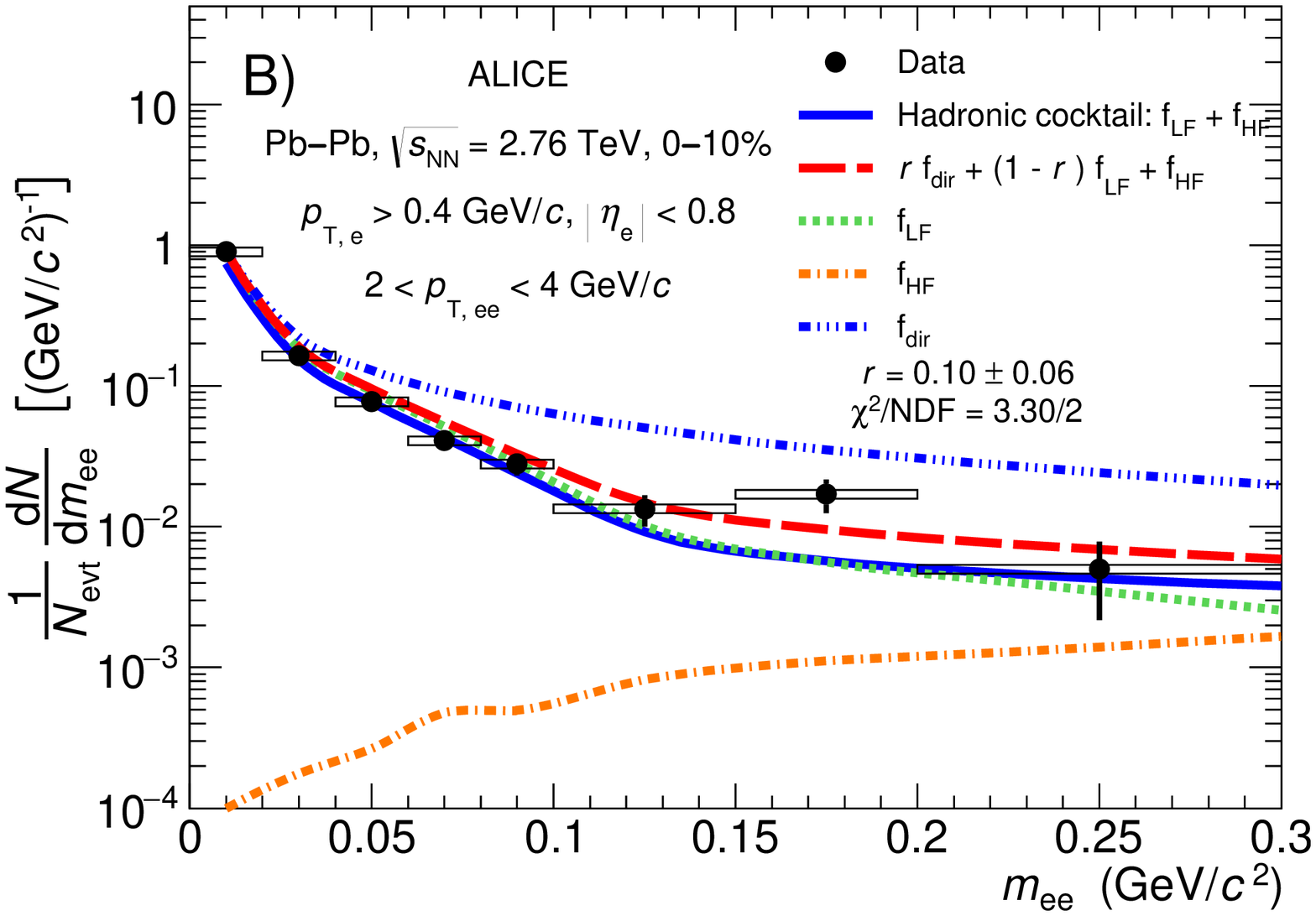}
         \caption{(colour online). Fit function (Eq.~\ref{eq:ThreeComponentFunc}), its individual components and dielectron invariant-mass spectra measured in central collisions for the transverse-momentum intervals $1<p_{\mathrm{T}, \mathrm{ee}}<2\ \mathrm{GeV}/\textit{c}$ (A) and $2<p_{\mathrm{T}, \mathrm{ee}}<4\ \mathrm{GeV}/\textit{c}$ (B). The hadronic cocktail is $f_{\mathrm{LF}} (m_{\mathrm{ee}}) + f_{\mathrm{HF}} (m_{\mathrm{ee}})$. }
 \label{fig:dielectronSpectrumAndFit}
 \end{figure}

The fit function, its individual components, the $\chi^{2}$ per degree of freedom of the fit, and the measured dielectron spectra in the transverse-momentum intervals $1<p_{\mathrm{T}, \mathrm{ee}}<2\ \mathrm{GeV}/\textit{c}$ and $2<p_{\mathrm{T}, \mathrm{ee}}<4\ \mathrm{GeV}/\textit{c}$ are shown in Fig.$~\ref{fig:dielectronSpectrumAndFit}$. 

For the calculation of the systematic uncertainties, the contributions from data, the hadronic cocktail components and the normalization range are considered separately. The systematic uncertainties from data are obtained by considering the variation of the virtual direct photon measurement due to a coherent shift of all data points by their systematic uncertainties, while the hadronic cocktail and the normalization range remained unaltered. 
The systematic uncertainties for the light-flavor and heavy-flavor components of the hadronic cocktail are calculated similarly, while the contribution from the normalization is calculated by considering the variations of the measurement corresponding to the following normalization ranges: $0 < m_{\mathrm{ee}} < 20\ \mathrm{MeV}/\textit{c}^{2}$, $0 < m_{\mathrm{ee}} < 40\ \mathrm{MeV}/\textit{c}^{2}$ and $0 < m_{\mathrm{ee}} < 60\ \mathrm{MeV}/\textit{c}^{2}$.
The systematic uncertainties on the fraction of virtual direct photons are summarized in Table~\ref{tab:SystematicUncertaintiesVirtualDirectPhotons}. The total systematic uncertainties are obtained by summing in quadrature all individual contributions.

\begin{table}[!hbt]
\centering
\renewcommand{\arraystretch}{1.6}
  \begin{tabular}{lcP{5cm}lcP{5cm}lcP{5cm}l}
\hline
  Source  &  $1<p_{\mathrm{T}, \mathrm{ee}}<2\ \mathrm{GeV}/\textit{c}$   &   $2<p_{\mathrm{T}, \mathrm{ee}}<4\ \mathrm{GeV}/\textit{c}$    \\
\hline
   Data     								         &  0.0014     &  0.0024   \\  \hline
   Light-flavor cocktail         					         &  0.0223     &  0.0193    \\  \hline
   Heavy-flavor cocktail       					         &  0.0123     &  0.0197    \\  \hline
   Normalization range    						&  0.0036     &  0.0215     \\  \hline
   Total    									&  0.0256     &  0.0314    \\   \hline
\end{tabular}
\caption{Summary of systematic uncertainties on the fraction of virtual direct photons. }
\label{tab:SystematicUncertaintiesVirtualDirectPhotons}
\end{table}

The result of the virtual direct photon measurement, including statistical and systematic uncertainties, is:
\newline
\newline
\centerline{$r=0.02 \pm 0.04 (\mathrm{stat.}) \pm 0.03 (\mathrm{syst.})$ for $1<p_{\mathrm{T}, \mathrm{ee}}<2\ \mathrm{GeV}/\textit{c}$,}
\newline 
\centerline{$r=0.10 \pm 0.06 (\mathrm{stat.}) \pm 0.03 (\mathrm{syst.})$ for $2<p_{\mathrm{T}, \mathrm{ee}}<4\ \mathrm{GeV}/\textit{c}$.}
\newline
\newline
Considering that every source of real photons is also a source of virtual photons, the fraction of real direct photons is expected to be equal to that of virtual direct photons in the zero-mass limit:

\begin{equation} 
\frac{(\gamma^{*})^{\mathrm{dir}}}{(\gamma^{*})^{\mathrm{incl}}} \xrightarrow[m_{\mathrm{ee}} \to 0\ \mathrm{GeV}/\textit{c}^{2}]{} \frac{\gamma^{\mathrm{dir}}}{\gamma^{\mathrm{incl}}}
\end{equation}

\begin{figure}[!hbt]
         \centering
                 \includegraphics[height=7cm,width=10cm]{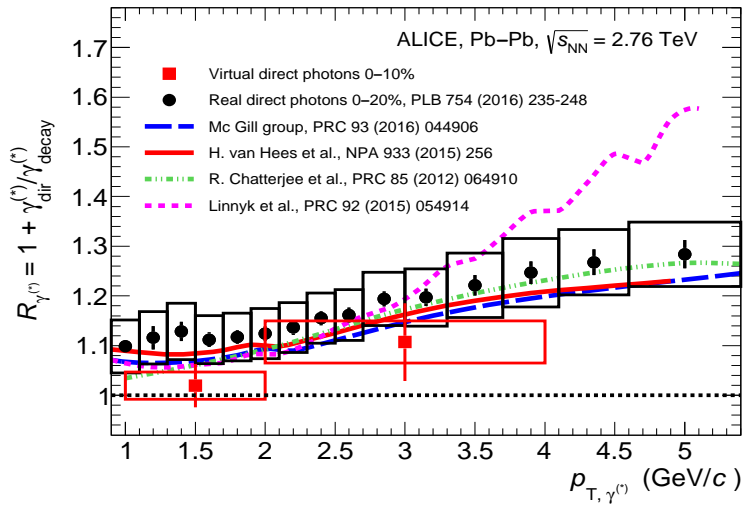}
         \caption{(colour online). Comparison between the direct photon ratio $R_{\gamma^{(*)}}=1+\gamma^{(*)}_{\mathrm{dir}}/\gamma^{(*)}_{\mathrm{decay}}$ measured using virtual photons in the centrality range 0$-$10$\%$ to that obtained using real photons in the centrality range 0$-$20$\%$ and several theoretical model calculations (see text for references). The real photon measurement is taken from \cite{directPhotonsPbPbALICE}. The statistical and systematic uncertainties of the data are represented by vertical bars and boxes. }
 \label{fig:VirtualDirectPhotonFraction}
 \end{figure}

In Fig.~\ref{fig:VirtualDirectPhotonFraction}, the fraction of virtual direct photons measured in Pb--Pb collisions at $\mathbf{\sqrt{{\textit{s}}_{\mathrm{NN}}}}$ = 2.76 TeV in the centrality range 0$-$10$\%$ is compared to the real direct photon measurement at the same center-of-mass energy in the centrality range 0$-$20$\%$ \cite{directPhotonsPbPbALICE} \footnote{The real direct photon measurement was performed using the 2010 data set, for which no trigger on central collisions was used. The measurement was performed in a wider centrality interval compared to that used in the analysis presented in this paper in order to improve the statistical precision of the measurement.} and to several theoretical model calculations related to the centrality range 0$-$10$\%$. The equivalent direct photon ratio for virtual photons is obtained as $R_{\gamma} = 1 / (1-r)$. 
In all the models, the contribution from prompt photons is calculated using pQCD, while the thermal component is obtained by integrating the static emission rate of thermal photons over the space-time evolution of the system assuming that a Quark-Gluon Plasma is formed in heavy-ion collisions. The main difference between these models is the description of the system evolution which results in different thermal photon contributions.
\newline
In the approach by Paquet $\textit{et al.}$ \cite{PaquetCalculation}, the system evolution follows a 2+1D hydrodynamical model, described in \cite{bulkViscosityHydroModel}. The initial conditions of the collision are obtained using the IP-Glasma approach \cite{fluctuatingGlasma}, with the formation time of the plasma set at $\tau_{0}$ = 0.4 fm/$\textit{c}$. Shear and bulk viscosity are also taken into account. In the hadronic phase, following the hydrodynamic expansion, the particle interactions are modeled by the Ultrarelativistic Quantum Molecular Dynamics (UrQMD) model \cite{UrQMD}. The photon emission rates from \cite{hadronicProductionThermalPhotons,universalParametrizationThermalPhotonRates,thermalPhotonEmissionFromPiRhoOmegaSystem} are adopted, including a viscous correction for each emission channel.
\newline
In van Hees $\textit{et al.}$ \cite{vanHees} calculations, a lattice-QCD equation of state (EoS) is adopted for the QGP phase and it is connected with a hadron gas at the freeze-out. Initial radial and elliptic flow effects are also included. The photon emission rates are taken from \cite{hadronicProductionThermalPhotons}. 
\newline
The Chatterjee $\textit{et al.}$ \cite{Chatterjee} model uses event-by-event hydrodynamics to represent the initial inhomogeneities of the energy density profile. The initial formation time is $\tau_{0}$ = 0.4 fm/$\textit{c}$. The 2+1D ideal hydrodynamic evolution is assumed to have longitudinal boost invariance and is solved with the SHASTA algorithm \cite{SHASTA}. The equation of state is from \cite{quarkMassThresholdsQCDthermodynamics}, while the emission rates are taken from \cite{EmissionRateQGP} for the QGP phase and from \cite{hadronicProductionThermalPhotons} for the hadronic phase. The total thermal photon emission rates are calculated integrating over the full fireball space-time evolution. 
\newline
The Linnyk $\textit{et al.}$ \cite{hadronicPartonicSourcesDirectPhotons,effectiveQCDTransportDescriptionDileptons} model uses the off-shell transport approach PHSD \cite{PHSDoffShellTransport,PHSDRHICEnergies} to give a microscopic description of the collision evolution. Both partonic and hadronic interactions are considered as sources of photons. The latter involves the production of photons from meson--meson or meson--baryon binary collisions or bremsstrahlung radiation as well as the production of photons in hadronic decays. In addition, vector meson and nucleon interactions and the $\Delta$ resonance decay are also considered. The prompt photon component is the same given in \cite{PaquetCalculation}.

The direct photon ratio measured in the centrality range 0$-$10$\%$ is expected to be larger than that measured in the centrality range 0$-$20$\%$ due to the larger average energy density and temperature in the most central Pb--Pb collisions. The data instead show that the direct photon ratio measured in the centrality range 0$-$10$\%$ is at the lower edge of that measured in 0-20$\%$, remaining however consistent with it within the statistical and systematic uncertainties. 
The predictions from the different theoretical model calculations are consistent with both measurements within the experimental uncertainties in the entire transverse momentum range, except for Linnyk $\textit{et al.}$ in which the contribution from prompt photons, which is dominant at high $p_{\mathrm{T}, \gamma^{(*)}}$, is overestimated.
\newline
The statistical significance of the virtual direct photon measurements, defined as the ratio between the measured value and its statistical uncertainty, is $r/\Delta r_{\mathrm{stat}} = 0.45$ for $1<p_{\mathrm{T}, \mathrm{ee}}<2\ \mathrm{GeV}/\textit{c}$ and $r/\Delta r_{\mathrm{stat}} = 1.51$ for $2<p_{\mathrm{T}, \mathrm{ee}}<4\ \mathrm{GeV}/\textit{c}$. 
Given the statistical significance smaller than 3, these measurements are consistent with zero within their statistical uncertainties. In consequence, an upper limit at 90$\%$ C.L. on the virtual direct photon production is estimated in both transverse-momentum intervals based on the Feldman and Cousins methodology \cite{FeldmanCousins}. Gaussian distributions are assumed for statistical and systematic uncertainties, that are treated independently and summed quadratically in the estimation of these confidence ranges. The estimated upper limits on the virtual direct photon ratios are 0.10 for $1<p_{\mathrm{T}, \mathrm{ee}}<2\ \mathrm{GeV}/\textit{c}$ and 0.22 for $2<p_{\mathrm{T}, \mathrm{ee}}<4\ \mathrm{GeV}/\textit{c}$.
\newline
The measurement of the fraction of virtual direct photon can be compared to the measurements from PHENIX and STAR in Au--Au collisions at $\mathbf{\sqrt{{\textit{s}}_{\mathrm{NN}}}}=200\ \mathrm{GeV}$ at RHIC \cite{dielectronProductionAuAuPHENIXnew,dielectronsAuAuSTAR}. The fraction of direct photons from the initial hard parton-parton scattering (prompt photons) is expected to decrease with increasing center-of-mass energy as the prompt direct photon cross section rises slower than the hadron cross section with collision energy \cite{dNchdEtaALICE,pQCDdirectPhotons}. In pp collisions, the fraction of prompt direct photons at the LHC energy is indeed lower than at RHIC (around 1$-$2$\%\ \mathrm{vs.}\ 3\%$) \cite{DirectPhotonsProtonProtonALICE}. The transverse momentum region $1<p_{\mathrm{T}, \mathrm{ee}}<4\ \mathrm{GeV}/\textit{c}$, where the fraction of virtual direct photons is measured, contains a significant contribution from thermal photons. 
The production yield of thermal photons at LHC energies is expected to be larger than at RHIC due to the higher initial temperature of the system, its larger size and longer lifetime. 
This scenario is confirmed by the fact that the fraction of virtual direct photons measured at RHIC in the same transverse momentum range, going from around $10\%$ at $p_{\mathrm{T}} \approx 1\ \mathrm{GeV}/\textit{c}$ to 20$-$30$\%$ at $p_{\mathrm{T}} \approx 4-5\ \mathrm{GeV}/\textit{c}$, is similar to the measurement presented in this paper.

\section{Summary and outlook}
\label{sec:Conclusions}

The first measurement of the dielectron invariant-mass spectrum in central (0$-$10$\%$) Pb--Pb collisions at $\mathbf{\sqrt{{\textit{s}}_{\mathrm{NN}}}}$ = 2.76 TeV was presented. 
A significant effort was made in the description of the hadron decay background and in the signal extraction for this extremely challenging measurement. 
An innovative technique, based on a single-track rejection, was used to suppress the contribution to the combinatorial background of electron-positron pairs produced by photon conversion in the detector material that is now extensively used for other dielectron measurements. 
The cocktail of known hadronic sources is consistent with the dielectron spectrum over the measured invariant-mass range within the statistical and systematic uncertainties. 
The data to cocktail ratio measured in the invariant-mass range $0.15 < m_{\mathrm{ee}}<0.7\ \mathrm{GeV}/\textit{c}^{2}$, excluding the contribution from the vacuum $\rho^{0}$, is $1.40 \pm 0.28\ (\mathrm{stat.}) \pm 0.08\ (\mathrm{syst.}) \pm 0.27\ (\mathrm{cocktail})$. 
The relatively low number of events collected during the Pb--Pb data taking period of 2011 and the limited knowledge of the charm contribution, which represents the dominant source of dielectrons with $m_{\mathrm{ee}}>0.4\ \mathrm{GeV}/\textit{c}^{2}$, reduce the sensitivity to a thermal signal in the low-mass region.
The dielectron spectrum is compared to two theoretical models which include the contributions of thermal dielectrons from partonic and hadronic phases and assume a broadening of the electromagnetic spectral function of the $\rho^{0}$ meson. Both models are consistent with the data within the uncertainties.
\newline
The effect of the interactions between charm quarks and other partons in the medium is simulated assuming random correlations between dielectrons from charm decays using PYTHIA simulations. The limited precision of the dielectron measurement in the invariant-mass region dominated by heavy-flavor decays ($m_{\phi} < m_{\mathrm{ee}} <m_{\mathrm{J}/\psi}$) prevents any conclusion on shadowing and energy loss effects on the dielectron spectrum. 
\newline
The fraction of virtual direct photons over inclusive virtual photons was measured in the invariant-mass range $100 < m_{\mathrm{ee}}<300\ \mathrm{MeV}/\textit{c}^{2}$ for $1<p_{\mathrm{T}, \mathrm{ee}}<2\ \mathrm{GeV}/\textit{c}$ and $2<p_{\mathrm{T}, \mathrm{ee}}<4\ \mathrm{GeV}/\textit{c}$. The measured fraction of virtual direct photons is at the lower edge of the real direct photon measurement from ALICE \cite{directPhotonsPbPbALICE}, remaining consistent with it within the experimental uncertainties. The virtual photon measurement is also consistent with the expectations from previous dielectron measurements at RHIC in Au--Au collisions at $\mathbf{\sqrt{{\textit{s}}_{\mathrm{NN}}}}=200\ \mathrm{GeV}$ \cite{dielectronProductionAuAuPHENIXnew,dielectronsAuAuSTAR}. 
\newline
A more precise dielectron measurement is expected with the new Pb--Pb data at $\mathbf{\sqrt{{\textit{s}}_{\mathrm{NN}}}}$ = 5.02 TeV of the LHC Run 2, while a significant improvement is expected after the ALICE upgrade, where the number of events is expected to increase by a factor 100. 
The simulations of the detector performance indicate that a detailed study of the in-medium properties of the $\rho^{0}$ meson will be done from the analysis of the excess in the low-mass region and the QGP temperature will be measured with a precision of 10$-$20$\%$ from an exponential fit to the intermediate mass region, where the contribution from charm decays is expected to be strongly suppressed by the improved secondary vertex resolution of the upgraded detectors \cite{ALICEUpgrade}.

\newenvironment{acknowledgement}{\relax}{\relax}
\begin{acknowledgement}
\section*{Acknowledgements}

The ALICE Collaboration would like to thank all its engineers and technicians for their invaluable contributions to the construction of the experiment and the CERN accelerator teams for the outstanding performance of the LHC complex.
The ALICE Collaboration gratefully acknowledges the resources and support provided by all Grid centres and the Worldwide LHC Computing Grid (WLCG) collaboration.
The ALICE Collaboration acknowledges the following funding agencies for their support in building and running the ALICE detector:
A. I. Alikhanyan National Science Laboratory (Yerevan Physics Institute) Foundation (ANSL), State Committee of Science and World Federation of Scientists (WFS), Armenia;
Austrian Academy of Sciences and Nationalstiftung f\"{u}r Forschung, Technologie und Entwicklung, Austria;
Ministry of Communications and High Technologies, National Nuclear Research Center, Azerbaijan;
Conselho Nacional de Desenvolvimento Cient\'{\i}fico e Tecnol\'{o}gico (CNPq), Universidade Federal do Rio Grande do Sul (UFRGS), Financiadora de Estudos e Projetos (Finep) and Funda\c{c}\~{a}o de Amparo \`{a} Pesquisa do Estado de S\~{a}o Paulo (FAPESP), Brazil;
Ministry of Science \& Technology of China (MSTC), National Natural Science Foundation of China (NSFC) and Ministry of Education of China (MOEC) , China;
Ministry of Science and Education, Croatia;
Ministry of Education, Youth and Sports of the Czech Republic, Czech Republic;
The Danish Council for Independent Research | Natural Sciences, the Carlsberg Foundation and Danish National Research Foundation (DNRF), Denmark;
Helsinki Institute of Physics (HIP), Finland;
Commissariat \`{a} l'Energie Atomique (CEA) and Institut National de Physique Nucl\'{e}aire et de Physique des Particules (IN2P3) and Centre National de la Recherche Scientifique (CNRS), France;
Bundesministerium f\"{u}r Bildung, Wissenschaft, Forschung und Technologie (BMBF) and GSI Helmholtzzentrum f\"{u}r Schwerionenforschung GmbH, Germany;
General Secretariat for Research and Technology, Ministry of Education, Research and Religions, Greece;
National Research, Development and Innovation Office, Hungary;
Department of Atomic Energy Government of India (DAE), Department of Science and Technology, Government of India (DST), University Grants Commission, Government of India (UGC) and Council of Scientific and Industrial Research (CSIR), India;
Indonesian Institute of Science, Indonesia;
Centro Fermi - Museo Storico della Fisica e Centro Studi e Ricerche Enrico Fermi and Istituto Nazionale di Fisica Nucleare (INFN), Italy;
Institute for Innovative Science and Technology , Nagasaki Institute of Applied Science (IIST), Japan Society for the Promotion of Science (JSPS) KAKENHI and Japanese Ministry of Education, Culture, Sports, Science and Technology (MEXT), Japan;
Consejo Nacional de Ciencia (CONACYT) y Tecnolog\'{i}a, through Fondo de Cooperaci\'{o}n Internacional en Ciencia y Tecnolog\'{i}a (FONCICYT) and Direcci\'{o}n General de Asuntos del Personal Academico (DGAPA), Mexico;
Nederlandse Organisatie voor Wetenschappelijk Onderzoek (NWO), Netherlands;
The Research Council of Norway, Norway;
Commission on Science and Technology for Sustainable Development in the South (COMSATS), Pakistan;
Pontificia Universidad Cat\'{o}lica del Per\'{u}, Peru;
Ministry of Science and Higher Education and National Science Centre, Poland;
Korea Institute of Science and Technology Information and National Research Foundation of Korea (NRF), Republic of Korea;
Ministry of Education and Scientific Research, Institute of Atomic Physics and Romanian National Agency for Science, Technology and Innovation, Romania;
Joint Institute for Nuclear Research (JINR), Ministry of Education and Science of the Russian Federation and National Research Centre Kurchatov Institute, Russia;
Ministry of Education, Science, Research and Sport of the Slovak Republic, Slovakia;
National Research Foundation of South Africa, South Africa;
Centro de Aplicaciones Tecnol\'{o}gicas y Desarrollo Nuclear (CEADEN), Cubaenerg\'{\i}a, Cuba and Centro de Investigaciones Energ\'{e}ticas, Medioambientales y Tecnol\'{o}gicas (CIEMAT), Spain;
Swedish Research Council (VR) and Knut \& Alice Wallenberg Foundation (KAW), Sweden;
European Organization for Nuclear Research, Switzerland;
National Science and Technology Development Agency (NSDTA), Suranaree University of Technology (SUT) and Office of the Higher Education Commission under NRU project of Thailand, Thailand;
Turkish Atomic Energy Agency (TAEK), Turkey;
National Academy of  Sciences of Ukraine, Ukraine;
Science and Technology Facilities Council (STFC), United Kingdom;
National Science Foundation of the United States of America (NSF) and United States Department of Energy, Office of Nuclear Physics (DOE NP), United States of America.    
\end{acknowledgement}

\bibliographystyle{utphys}   
\bibliography{bibliography}  
\newpage
\appendix
\section{The ALICE Collaboration}
\label{app:collab}

\begingroup
\small
\begin{flushleft}
S.~Acharya\Irefn{org139}\And 
F.T.-.~Acosta\Irefn{org20}\And 
D.~Adamov\'{a}\Irefn{org93}\And 
J.~Adolfsson\Irefn{org80}\And 
M.M.~Aggarwal\Irefn{org98}\And 
G.~Aglieri Rinella\Irefn{org34}\And 
M.~Agnello\Irefn{org31}\And 
N.~Agrawal\Irefn{org48}\And 
Z.~Ahammed\Irefn{org139}\And 
S.U.~Ahn\Irefn{org76}\And 
S.~Aiola\Irefn{org144}
% ...
\And
\renewcommand\labelenumi{\textsuperscript{\theenumi}~}

\section*{Affiliation notes}
\renewcommand\theenumi{\roman{enumi}}
\begin{Authlist}
\item \Adef{org*}Deceased
\item \Adef{orgI}Dipartimento DET del Politecnico di Torino, Turin, Italy
\item \Adef{orgII}M.V. Lomonosov Moscow State University, D.V. Skobeltsyn Institute of Nuclear, Physics, Moscow, Russia
\item \Adef{orgIII}Department of Applied Physics, Aligarh Muslim University, Aligarh, India
\item \Adef{orgIV}Institute of Theoretical Physics, University of Wroclaw, Poland
\end{Authlist}

\section*{Collaboration Institutes}
\renewcommand\theenumi{\arabic{enumi}~}
\begin{Authlist}
\item \Idef{org1}A.I. Alikhanyan National Science Laboratory (Yerevan Physics Institute) Foundation, Yerevan, Armenia
\item \Idef{org2}Bogolyubov Institute for Theoretical Physics, National Academy of Sciences of Ukraine, Kiev, Ukraine
\item \Idef{org3}Bose Institute, Department of Physics  and Centre for Astroparticle Physics and Space Science (CAPSS), Kolkata, India
\item \Idef{org4}Budker Institute for Nuclear Physics, Novosibirsk, Russia
\item \Idef{org5}California Polytechnic State University, San Luis Obispo, California, United States
\item \Idef{org6}Central China Normal University, Wuhan, China
\item \Idef{org7}Centre de Calcul de l'IN2P3, Villeurbanne, Lyon, France
\item \Idef{org8}Centro de Aplicaciones Tecnol\'{o}gicas y Desarrollo Nuclear (CEADEN), Havana, Cuba
\item \Idef{org9}Centro de Investigaci\'{o}n y de Estudios Avanzados (CINVESTAV), Mexico City and M\'{e}rida, Mexico
\item \Idef{org10}Centro Fermi - Museo Storico della Fisica e Centro Studi e Ricerche ``Enrico Fermi', Rome, Italy
\item \Idef{org11}Chicago State University, Chicago, Illinois, United States
\item \Idef{org12}China Institute of Atomic Energy, Beijing, China
\item \Idef{org13}Chonbuk National University, Jeonju, Republic of Korea
\item \Idef{org14}Comenius University Bratislava, Faculty of Mathematics, Physics and Informatics, Bratislava, Slovakia
\item \Idef{org15}COMSATS Institute of Information Technology (CIIT), Islamabad, Pakistan
\item \Idef{org16}Creighton University, Omaha, Nebraska, United States
\item \Idef{org17}Department of Physics, Aligarh Muslim University, Aligarh, India
\item \Idef{org18}Department of Physics, Pusan National University, Pusan, Republic of Korea
\item \Idef{org19}Department of Physics, Sejong University, Seoul, Republic of Korea
\item \Idef{org20}Department of Physics, University of California, Berkeley, California, United States
\item \Idef{org21}Department of Physics, University of Oslo, Oslo, Norway
\item \Idef{org22}Department of Physics and Technology, University of Bergen, Bergen, Norway
\item \Idef{org23}Dipartimento di Fisica dell'Universit\`{a} 'La Sapienza' and Sezione INFN, Rome, Italy
\item \Idef{org24}Dipartimento di Fisica dell'Universit\`{a} and Sezione INFN, Cagliari, Italy
\item \Idef{org25}Dipartimento di Fisica dell'Universit\`{a} and Sezione INFN, Trieste, Italy
\item \Idef{org26}Dipartimento di Fisica dell'Universit\`{a} and Sezione INFN, Turin, Italy
\item \Idef{org27}Dipartimento di Fisica e Astronomia dell'Universit\`{a} and Sezione INFN, Bologna, Italy
\item \Idef{org28}Dipartimento di Fisica e Astronomia dell'Universit\`{a} and Sezione INFN, Catania, Italy
\item \Idef{org29}Dipartimento di Fisica e Astronomia dell'Universit\`{a} and Sezione INFN, Padova, Italy
\item \Idef{org30}Dipartimento di Fisica `E.R.~Caianiello' dell'Universit\`{a} and Gruppo Collegato INFN, Salerno, Italy
\item \Idef{org31}Dipartimento DISAT del Politecnico and Sezione INFN, Turin, Italy
\item \Idef{org32}Dipartimento di Scienze e Innovazione Tecnologica dell'Universit\`{a} del Piemonte Orientale and INFN Sezione di Torino, Alessandria, Italy
\item \Idef{org33}Dipartimento Interateneo di Fisica `M.~Merlin' and Sezione INFN, Bari, Italy
\item \Idef{org34}European Organization for Nuclear Research (CERN), Geneva, Switzerland
\item \Idef{org35}Faculty of Electrical Engineering, Mechanical Engineering and Naval Architecture, University of Split, Split, Croatia
\item \Idef{org36}Faculty of Engineering and Science, Western Norway University of Applied Sciences, Bergen, Norway
\item \Idef{org37}Faculty of Nuclear Sciences and Physical Engineering, Czech Technical University in Prague, Prague, Czech Republic
\item \Idef{org38}Faculty of Science, P.J.~\v{S}af\'{a}rik University, Ko\v{s}ice, Slovakia
\item \Idef{org39}Frankfurt Institute for Advanced Studies, Johann Wolfgang Goethe-Universit\"{a}t Frankfurt, Frankfurt, Germany
\item \Idef{org40}Gangneung-Wonju National University, Gangneung, Republic of Korea
\item \Idef{org41}Gauhati University, Department of Physics, Guwahati, India
\item \Idef{org42}Helmholtz-Institut f\"{u}r Strahlen- und Kernphysik, Rheinische Friedrich-Wilhelms-Universit\"{a}t Bonn, Bonn, Germany
\item \Idef{org43}Helsinki Institute of Physics (HIP), Helsinki, Finland
\item \Idef{org44}High Energy Physics Group,  Universidad Aut\'{o}noma de Puebla, Puebla, Mexico
\item \Idef{org45}Hiroshima University, Hiroshima, Japan
\item \Idef{org46}Hochschule Worms, Zentrum  f\"{u}r Technologietransfer und Telekommunikation (ZTT), Worms, Germany
\item \Idef{org47}Horia Hulubei National Institute of Physics and Nuclear Engineering, Bucharest, Romania
\item \Idef{org48}Indian Institute of Technology Bombay (IIT), Mumbai, India
\item \Idef{org49}Indian Institute of Technology Indore, Indore, India
\item \Idef{org50}Indonesian Institute of Sciences, Jakarta, Indonesia
\item \Idef{org51}INFN, Laboratori Nazionali di Frascati, Frascati, Italy
\item \Idef{org52}INFN, Sezione di Bari, Bari, Italy
\item \Idef{org53}INFN, Sezione di Bologna, Bologna, Italy
\item \Idef{org54}INFN, Sezione di Cagliari, Cagliari, Italy
\item \Idef{org55}INFN, Sezione di Catania, Catania, Italy
\item \Idef{org56}INFN, Sezione di Padova, Padova, Italy
\item \Idef{org57}INFN, Sezione di Roma, Rome, Italy
\item \Idef{org58}INFN, Sezione di Torino, Turin, Italy
\item \Idef{org59}INFN, Sezione di Trieste, Trieste, Italy
\item \Idef{org60}Inha University, Incheon, Republic of Korea
\item \Idef{org61}Institut de Physique Nucl\'{e}aire d'Orsay (IPNO), Institut National de Physique Nucl\'{e}aire et de Physique des Particules (IN2P3/CNRS), Universit\'{e} de Paris-Sud, Universit\'{e} Paris-Saclay, Orsay, France
\item \Idef{org62}Institute for Nuclear Research, Academy of Sciences, Moscow, Russia
\item \Idef{org63}Institute for Subatomic Physics, Utrecht University/Nikhef, Utrecht, Netherlands
\item \Idef{org64}Institute for Theoretical and Experimental Physics, Moscow, Russia
\item \Idef{org65}Institute of Experimental Physics, Slovak Academy of Sciences, Ko\v{s}ice, Slovakia
\item \Idef{org66}Institute of Physics, Homi Bhabha National Institute, Bhubaneswar, India
\item \Idef{org67}Institute of Physics of the Czech Academy of Sciences, Prague, Czech Republic
\item \Idef{org68}Institute of Space Science (ISS), Bucharest, Romania
\item \Idef{org69}Institut f\"{u}r Kernphysik, Johann Wolfgang Goethe-Universit\"{a}t Frankfurt, Frankfurt, Germany
\item \Idef{org70}Instituto de Ciencias Nucleares, Universidad Nacional Aut\'{o}noma de M\'{e}xico, Mexico City, Mexico
\item \Idef{org71}Instituto de F\'{i}sica, Universidade Federal do Rio Grande do Sul (UFRGS), Porto Alegre, Brazil
\item \Idef{org72}Instituto de F\'{\i}sica, Universidad Nacional Aut\'{o}noma de M\'{e}xico, Mexico City, Mexico
\item \Idef{org73}iThemba LABS, National Research Foundation, Somerset West, South Africa
\item \Idef{org74}Johann-Wolfgang-Goethe Universit\"{a}t Frankfurt Institut f\"{u}r Informatik, Fachbereich Informatik und Mathematik, Frankfurt, Germany
\item \Idef{org75}Joint Institute for Nuclear Research (JINR), Dubna, Russia
\item \Idef{org76}Korea Institute of Science and Technology Information, Daejeon, Republic of Korea
\item \Idef{org77}KTO Karatay University, Konya, Turkey
\item \Idef{org78}Laboratoire de Physique Subatomique et de Cosmologie, Universit\'{e} Grenoble-Alpes, CNRS-IN2P3, Grenoble, France
\item \Idef{org79}Lawrence Berkeley National Laboratory, Berkeley, California, United States
\item \Idef{org80}Lund University Department of Physics, Division of Particle Physics, Lund, Sweden
\item \Idef{org81}Nagasaki Institute of Applied Science, Nagasaki, Japan
\item \Idef{org82}Nara Women{'}s University (NWU), Nara, Japan
\item \Idef{org83}National and Kapodistrian University of Athens, School of Science, Department of Physics , Athens, Greece
\item \Idef{org84}National Centre for Nuclear Research, Warsaw, Poland
\item \Idef{org85}National Institute of Science Education and Research, Homi Bhabha National Institute, Jatni, India
\item \Idef{org86}National Nuclear Research Center, Baku, Azerbaijan
\item \Idef{org87}National Research Centre Kurchatov Institute, Moscow, Russia
\item \Idef{org88}Niels Bohr Institute, University of Copenhagen, Copenhagen, Denmark
\item \Idef{org89}Nikhef, National institute for subatomic physics, Amsterdam, Netherlands
\item \Idef{org90}NRC Kurchatov Institute IHEP, Protvino, Russia
\item \Idef{org91}NRNU Moscow Engineering Physics Institute, Moscow, Russia
\item \Idef{org92}Nuclear Physics Group, STFC Daresbury Laboratory, Daresbury, United Kingdom
\item \Idef{org93}Nuclear Physics Institute of the Czech Academy of Sciences, \v{R}e\v{z} u Prahy, Czech Republic
\item \Idef{org94}Oak Ridge National Laboratory, Oak Ridge, Tennessee, United States
\item \Idef{org95}Ohio State University, Columbus, Ohio, United States
\item \Idef{org96}Petersburg Nuclear Physics Institute, Gatchina, Russia
\item \Idef{org97}Physics department, Faculty of science, University of Zagreb, Zagreb, Croatia
\item \Idef{org98}Physics Department, Panjab University, Chandigarh, India
\item \Idef{org99}Physics Department, University of Jammu, Jammu, India
\item \Idef{org100}Physics Department, University of Rajasthan, Jaipur, India
\item \Idef{org101}Physikalisches Institut, Eberhard-Karls-Universit\"{a}t T\"{u}bingen, T\"{u}bingen, Germany
\item \Idef{org102}Physikalisches Institut, Ruprecht-Karls-Universit\"{a}t Heidelberg, Heidelberg, Germany
\item \Idef{org103}Physik Department, Technische Universit\"{a}t M\"{u}nchen, Munich, Germany
\item \Idef{org104}Research Division and ExtreMe Matter Institute EMMI, GSI Helmholtzzentrum f\"ur Schwerionenforschung GmbH, Darmstadt, Germany
\item \Idef{org105}Rudjer Bo\v{s}kovi\'{c} Institute, Zagreb, Croatia
\item \Idef{org106}Russian Federal Nuclear Center (VNIIEF), Sarov, Russia
\item \Idef{org107}Saha Institute of Nuclear Physics, Homi Bhabha National Institute, Kolkata, India
\item \Idef{org108}School of Physics and Astronomy, University of Birmingham, Birmingham, United Kingdom
\item \Idef{org109}Secci\'{o}n F\'{\i}sica, Departamento de Ciencias, Pontificia Universidad Cat\'{o}lica del Per\'{u}, Lima, Peru
\item \Idef{org110}Shanghai Institute of Applied Physics, Shanghai, China
\item \Idef{org111}St. Petersburg State University, St. Petersburg, Russia
\item \Idef{org112}Stefan Meyer Institut f\"{u}r Subatomare Physik (SMI), Vienna, Austria
\item \Idef{org113}SUBATECH, IMT Atlantique, Universit\'{e} de Nantes, CNRS-IN2P3, Nantes, France
\item \Idef{org114}Suranaree University of Technology, Nakhon Ratchasima, Thailand
\item \Idef{org115}Technical University of Ko\v{s}ice, Ko\v{s}ice, Slovakia
\item \Idef{org116}Technische Universit\"{a}t M\"{u}nchen, Excellence Cluster 'Universe', Munich, Germany
\item \Idef{org117}The Henryk Niewodniczanski Institute of Nuclear Physics, Polish Academy of Sciences, Cracow, Poland
\item \Idef{org118}The University of Texas at Austin, Austin, Texas, United States
\item \Idef{org119}Universidad Aut\'{o}noma de Sinaloa, Culiac\'{a}n, Mexico
\item \Idef{org120}Universidade de S\~{a}o Paulo (USP), S\~{a}o Paulo, Brazil
\item \Idef{org121}Universidade Estadual de Campinas (UNICAMP), Campinas, Brazil
\item \Idef{org122}Universidade Federal do ABC, Santo Andre, Brazil
\item \Idef{org123}University College of Southeast Norway, Tonsberg, Norway
\item \Idef{org124}University of Cape Town, Cape Town, South Africa
\item \Idef{org125}University of Houston, Houston, Texas, United States
\item \Idef{org126}University of Jyv\"{a}skyl\"{a}, Jyv\"{a}skyl\"{a}, Finland
\item \Idef{org127}University of Liverpool, Liverpool, United Kingdom
\item \Idef{org128}University of Tennessee, Knoxville, Tennessee, United States
\item \Idef{org129}University of the Witwatersrand, Johannesburg, South Africa
\item \Idef{org130}University of Tokyo, Tokyo, Japan
\item \Idef{org131}University of Tsukuba, Tsukuba, Japan
\item \Idef{org132}Universit\'{e} Clermont Auvergne, CNRS/IN2P3, LPC, Clermont-Ferrand, France
\item \Idef{org133}Universit\'{e} de Lyon, Universit\'{e} Lyon 1, CNRS/IN2P3, IPN-Lyon, Villeurbanne, Lyon, France
\item \Idef{org134}Universit\'{e} de Strasbourg, CNRS, IPHC UMR 7178, F-67000 Strasbourg, France, Strasbourg, France
\item \Idef{org135} Universit\'{e} Paris-Saclay Centre d¿\'Etudes de Saclay (CEA), IRFU, Department de Physique Nucl\'{e}aire (DPhN), Saclay, France
\item \Idef{org136}Universit\`{a} degli Studi di Foggia, Foggia, Italy
\item \Idef{org137}Universit\`{a} degli Studi di Pavia, Pavia, Italy
\item \Idef{org138}Universit\`{a} di Brescia, Brescia, Italy
\item \Idef{org139}Variable Energy Cyclotron Centre, Homi Bhabha National Institute, Kolkata, India
\item \Idef{org140}Warsaw University of Technology, Warsaw, Poland
\item \Idef{org141}Wayne State University, Detroit, Michigan, United States
\item \Idef{org142}Westf\"{a}lische Wilhelms-Universit\"{a}t M\"{u}nster, Institut f\"{u}r Kernphysik, M\"{u}nster, Germany
\item \Idef{org143}Wigner Research Centre for Physics, Hungarian Academy of Sciences, Budapest, Hungary
\item \Idef{org144}Yale University, New Haven, Connecticut, United States
\item \Idef{org145}Yonsei University, Seoul, Republic of Korea
\end{Authlist}
\endgroup
\end{document}